# New Variable Compliance Method for Estimating Closure Stress and Fracture Compliance from DFIT data*

HanYi Wang and Mukul M. Sharma

Petroleum and Geosystem Engineering Department, The University of Texas at Austin, USA

## Abstract

Over the past two decades, Diagnostic Fracture Injection Tests (DFIT), which have also been referred to as Injection-Falloff Tests, Fracture Calibration Tests, Mini-Frac Tests in the literature, have evolved into a commonly used and reliable technique to evaluate reservoir properties, fracturing parameters and obtain in-situ stresses. Since the introduction of DFIT analysis based on G-function and its derivative, this method has become standard practice for quantifying minimum in-situ stress and leak-off coefficient. However, the pressure decline model that underlies the G-function plot makes two distinct and important assumptions: (1) leak-off is not pressure-dependent and, (2) fracture stiffness (or compliance) is assumed to be constant during fracture closure. In this study, we first review Nolte's original G-function model and examine the assumptions inherent in the model. We then present a new global pressure transient model for pressure decline after shut-in which not only preserves the physics of unsteady-state reservoir flow behavior, elastic fracture mechanics and material balance, but also incorporates the gradual changes of fracture stiffness (or compliance) due to the contact of rough fracture walls during closure. Analysis of synthetic cases, along with field data are presented to demonstrate how the coupled effects of fracture geometry, fracture surface asperities, formation properties, pore pressure and wellbore storage can impact fracturing pressure decline and the estimation of minimum in-situ stress. It is shown that using Carter's leak-off is an oversimplification that leads to significant errors in the interpretation of DFIT data. Most importantly, this article reveals that previous methods of estimating minimum in-situ stress often lead to significant over or underestimates. Based on our modeling and simulation results, we propose a much more accurate and reliable method to estimate the minimum in-situ stress and fracture pressure dependent leak-off rate.

## 1. Introduction

Diagnostic fracture injection tests (DFIT) involve pumping a fluid (typically water), at a constant rate for a short period of time, creating a relatively small hydraulic fracture before the well is shut in. The pressure transient data after shut-in is analyzed to obtain in-situ stresses and reservoir properties. A typical pressure trend is qualitatively shown in **Fig.1**. Conventionally, DFIT analysis has focused on acquiring fracturing treatment design parameters, such as fluid efficiencies, leak-off coefficient and fracture closure pressure (which is normally interpreted as minimum in-situ stress). However, in recent years, DFIT analysis has been extended to obtain reservoir properties such as reservoir pore pressure, and permeability in unconventional reservoirs, where traditional pressure transient tests are impractical. The reservoir properties determined by DFIT are representative, because the created fracture can pass through near-wellbore damage zone and provide a large volume of investigation for true formation properties. In addition, the net pressure trends in a DFIT can be also used to infer the induced fracture complexity in different geological settings (Potocki 2012). Thoese valuable information obtained from DFITs provides key input parameters for modeling hydraulic fracture propagation (Wang 2015; Wang 2016; Wang et al. 2016), stimulation design (Ramurthy et al. 2011), development of reservoir models (Mirani et al. 2016; Loughry et al. 2015; Wang 2017) and post-fracture analysis (Fu et al. 2017). Without a realistic estimation of the fracturing parameters and reservoir properties, it would not be possible to optimize hydraulic fracture design and evaluate the economic viability of producing hydrocarbons from unconventional reservoirs.

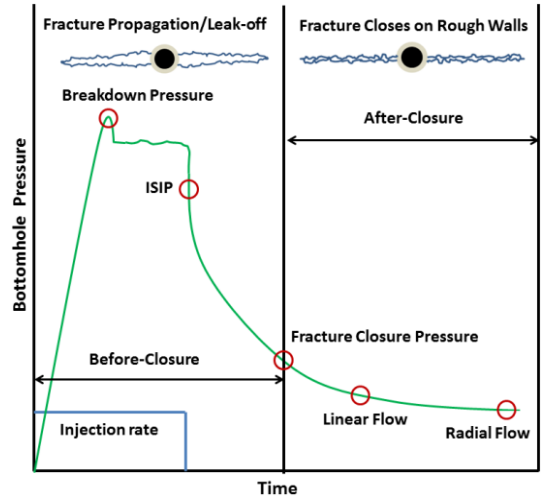

**Fig. 1 Diagram showing sequence of events observed in a DFIT**

The pressure decline data after shut-in can be divided into two distinct regions for analysis: before closure pressure data and after closure pressure data (separated by the fracture closure pressure). Analysis models are developed differently for these regimes. Before closure analysis (BCA) assumes normal leak-off behavior or uses the portion of normal leak-off data (a straight line on G function and square root of time plots) to estimate fracture geometry, leak-off coefficient or formation permeability (Mayerhofer et al. 1995; Nolte 1979 and1986; Valko ´and Economides 1999). A departure from the straight line indicates fracture closure (Castillo 1987). After closure analysis (ACA) models have been developed by several authors (Benelkadi and Tiab 2004; Chipperfield 2006; Nolte et al. 1997; Soliman et al. 2005; Soliman et al. 2010) to estimate pore pressure and infer formation permeability. Because both BCA and ACA only use some portion of the DFIT data, global consistency of interpretation can't be guaranteed, even if the individual analyses match a certain portion of the data well. Liu et al. (2016) and Marongiu-Porcu et al. (2014) attempted to bridge BCA and ACA on a log-log plot and the closure pressure is picked at the end of 3/2 slope of the pressure derivative. However, their work is still based on G-function model's assumptions and the 3/2 slope just arises from a spatial integration of Carter's leak-off (McClure 2017), so the end of 3/2 slope on log-log plot can't be used to justify fracture closure pressure.

The advent of fracturing pressure decline analysis was pioneered by the work of Nolte (1979 and1986). With the assumptions of power law fracture growth, negligible spurt loss, constant fracture surface area after shut-in and Carter's leak-off model (one-dimensional leak-off of fluid from a constant pressure boundary, the solution to the diffusivity equation predicts that the leak-off rate will scale with the inverse of the square root of time), a remarkably simple and useful equation for the pressure decline can be obtained:

$$P_f(\Delta t_D) = \text{ISIP} - \frac{\pi r_p C_L S_f \sqrt{t_p}}{2} G(\Delta t_D) \qquad (1)$$

Here, ISIP is the instantaneous shut-in pressure at the end of pumping, $P_f$ is the fracture pressure at dimensionless time $\Delta t_D$. $t_p$ is the total pumping time. $r_p$ is the productive fracture ratio, which is the ratio of fracture surface area that is subject to leak-off to the total fracture surface area. For low permeability, unconventional reservoirs, $r_p \approx 1$. To increase the readability of this article, $r_p$ will be assumed to be 1 for the rest of the discussion. $C_L$ is Carter's leak-off coefficient which is a constant. $S_f$ is the fracture stiffness, which can be calculated using **Table 1** for different fracture geometries. It is assumed that $S_f$ is a constant until the fracture closes instantaneously when the fluid pressure in the fracture reaches the closure pressure. Because fracture stiffness is the reciprocal of fracture compliance, they will be used interchangeably throughout this article.

| Fracture Geometry | PKN | KGD | Radial |
|---|---|---|---|
| $S_f$ | $\dfrac{2E'}{\pi h_f}$ | $\dfrac{E'}{\pi x_f}$ | $\dfrac{3\pi E'}{16 R_f}$ |

**Table 1-Fracture stiffness expressions for different fracture geometry models**

$E'$ is the plane strain Young's modulus and can be calculated using Young's Modulus, $E$ , and Poisson's Ratio, $\upsilon$ :

$$E' = \frac{E}{1 - \upsilon^2} \qquad (2)$$

The dimensionless time $\Delta t_D$ is defined by:

$$\Delta t_D = \frac{\Delta t}{t_e} = \frac{t - t_e}{t_e} \tag{3}$$

Where t is the generic time and $t_e$ is the time at the end of pumping. G-function is defined as

$$G(\Delta t_D) = \frac{4}{\pi}[g(\Delta t_D) - g(0)] \tag{4}$$

Where the g-function of time is approximated by,

$$g(\Delta t_D) = \begin{cases} (1 + \Delta t_D)\sin^{-1}(1 + \Delta t_D)^{-1/2} + \Delta t_D^{1/2} & \text{for low fluid efficiency} \\ \frac{4}{3}\left[(1 + \Delta t_D)^{1.5} - \Delta t_D^{1.5}\right] & \text{for high fluid efficiency} \end{cases} \tag{5}$$

From Eq.(1), we can infer that for normal leak-off behavior (constant Carter's leak-off coefficient, constant leak-off area and constant fracture stiffness during fracture closure), the pressure declines linearly with $G(\Delta t_D)$. Castillo (1987) used Nolte's G-function for modeling the pressure decline behavior and developed the straight-line plot of the G-function vs. pressure. The slope of this curve is used for the computation of the leak-off coefficient that is independent of pressure. Any departure from this straight-line is interpreted as closure of the fracture.

Unfortunately, plots of pressure versus G-function often yield curves with multiple points of inflection that have been attributed to abnormal leak-off behavior (such as pressure dependent leak-off, fracture height recession, closing of secondary transverse fractures, and fracture tip extension), which makes it difficult to interpret the changes in slope and identify fracture closure. So identification of fracture closure pressure and non-ideal behavior is usually done using plots of pressure and GdP/dG versus G-function (Barree and Mukherjee, 1996; Barree, 1998, Barree et al., 2014), where the closure is picked at the tangential point between a straight line that passes the origin and the GdP/dG curve. This prevailing method of determining minimum in-situ stress (although has been widely accepted, but has never been theoretically proved) will be classified and discussed as "tangent line method" in the following text. The fracturing pressure decline model that underlying G-function plot suffers from two distinct and important issues: (1) leak-off is not pressure-dependent, i.e. a constant pressure boundary is assumed (2) fracture compliance/stiffness is assumed to be constant during fracture closure. This is why G-function based models are only used for before closure analysis and are not capable of analyzing DFIT data from the end of pumping to days or even weeks after shut-in, which requires bridging both before and after closure data seamlessly.

McClure et al. (2016) modeled fracture closure behavior using a fully coupled numerical simulator and found that the "tangent line method" can severely underestimate closure pressure, and based on the simulation results, they proposed a "compliance method" for picking closure pressure on the G-function plot, where closure pressure is picked at the point where the fracture stiffness starts to increase. Zanganeh et al (2017) presented a cohesive zone fracture model to investigate fracture closure behavior, and their results also show that using the "tangent line method" or using the end of 3/2 slope on a log-log plot can underestimate closure pressure, and that the "compliance method" is a more reliable approach. However, fracture stiffness (or compliance) can change continuously during the fracture closure process, because the fracture does not close all at once, but rather closes on asperities progressively from its edges to the center (Wang and Sharma 2017; Wang et al 2017). So the mechanical closure pressure identified by "compliance method" can be larger than the minimum principal stress.

To better understanding how fracture stiffness, in-situ stress and pressure dependent leak-off impact the pressure decline signature, a global DFIT model that is able to simulate pressure decline under a single, coherent mathematic framework, and correctly capture the variable fracture compliance during closure is needed, and currently, no such model is available.  In this study, we present a generalization of the pressure transient model that couples the fluid pressure in the fracture with a pressure dependent leak-off rate and variable fracture compliance. Past models will be shown to be special cases for this general approach under certain simplifying assumptions..

## 2. Mathematical Formulation

### 2.1 The General Equations for a Global DFIT Model

The transient pressure response during fracture closure is derived using the following assumptions:

1. Reservoir is isotropic and homogeneous and contains a singlef slightly compressible fluid.

2. Reservoir permeability is low so that poroelastic effects caused by fluid leak-off are negligible

3. Fracture surface area subject to leak-off remains constant during and after closure. There are two stages of fracture closure: mechanical closure, where the fracture surfaces come into contact at asperities, and the other is hydraulic closure, where fracture pressure in the closed section is disconnected from the open section or wellbore pressure. In the following derivations, we assume that the mechanically closed fracture still retains hydraulic conductivity because of its residual fracture width that supported by asperities that caused by erosion or distortion of fracture walls.

4. Fracture has infinite conductivity after closure and pressure is uniformly distributed inside the fracture. This is the typical case in unconventional reservoirs. And in such case, the pressure distribution inside fracture can be considered as uniform during closure, as discussed by Koning et al. (1985)

5. The pressure disturbance caused by fracture propagation is negligible. This means that fluid leak-off during pumping is very small and the duration of injection is very short (typically 3-5 minutes) while the total shut-in time can be hours, days or even weeks.

In order to correctly capture fracturing pressure response during a DFIT, the pressure dependent leak-off at the fracture surface and the dynamic changes of fracture compliance during closure have to be accounted for. **Fig.2** illustrates one-dimensional leak-off into a semi-infinite formation.

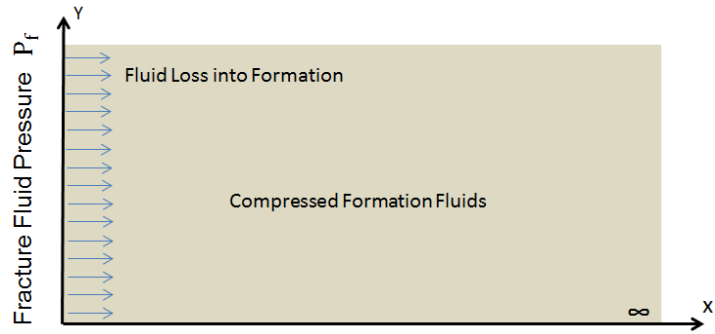

**Fig. 2 Illustration of one-dimensional leak-off**

Assuming linear Darcy flow and a slightly compressible, single phase fluid in the reservoir, the differential form of the mass balance can be written as:

$$\frac{\mu_f \phi c_t}{k} \frac{\partial P}{\partial t} = \frac{\partial^2 P}{\partial x^2} \qquad (6)$$

where $P$ is the pressure, k is formation permeability, $\mu_f$ is fluid viscosity, $\phi$ is formation porosity and $c_t$ is total formation compressibility. Inside the fracture, the average fracture width $\bar{w}_f$ and fracturing net pressure $P_{net}$ is related by the fracture stiffness:

$$P_{net} = S_f \bar{w}_f \qquad (7)$$

From a material balance perspective (fluid compressibility is negligible compared to that of the fracture), the rate of fluid leak-off into the formation, $q_f$ (one wing of the fracture), after shut-in equals the rate of shrinkage of fracture volume, $V_f$ (one wing of the fracture), as pressure declines:

$$q_f = -\frac{dV_f}{dt} \qquad (8)$$

And

$$\frac{dV_f}{dt} = \frac{dA_f \bar{w}_f}{dt} = \frac{A_f}{S_f} \frac{dP_{net}}{dt} = \frac{A_f}{S_f} \frac{dP_f}{dt} \qquad (9)$$

where $A_f$ is the fracture surface area of one face of one wing of the fracture. So Eq.(8) can be re-written as

$$q_f = -\frac{A_f}{S_f} \frac{dP_f}{dt} \qquad (10)$$

From the definition, we can conclude that the fracture stiffness (or compliance) is a representation of fracture compressibility per surface area. Higher stiffness (or lower compliance) implies that the fracture is less compressible. At the boundary of the fracture surface, both Darcy's flow and material balance have to be honored

$$q_f = -2A_f \frac{k}{\mu_f} \frac{dP}{dx} = -\frac{A_f}{S_f} \frac{dP}{dt} \text{ at x = 0} \tag{11}$$

Rearranging Eq. (11), we get the boundary condition at the fracture surface

$$\frac{k}{\mu_f} \frac{dP}{dx} = \frac{1}{2S_f} \frac{dP}{dt} \text{ at x = 0} \tag{12}$$

In the above derivation, it is assumed that the whole fracture surface area is subject to leak-off, which is the norm in unconventional reservoirs. If only a portion of fracture surface is considered permeable, then one needs to multiply the left side of Eq.(12) by the productive fracture ratio $r_p$

With initial condition (disregarding the pressure disturbance during the short injection period)

$$P = P_0 \text{ at t = 0, x > 0} \tag{13}$$

$$P = \text{ISIP at t = 0, x = 0} \tag{14}$$

where $P_0$ is the initial pore pressure.

The governing Eq.(6), plus the initial condition of Eq. (13)-(14) and boundary condition of Eq.(11) uniquely describes the pressure transient behavior during DFITs. In the above derivations, the influence of wellbore storage (WBS) is not included. For typical DFITs, untreated water is normally used as injection fluid. Since water is slightly compressible, the wellbore fluid will expand after shut-in with the declining pressure. Ignoring the early time fluid transient behavior along the wellbore after shut-in and assuming fracturing pressure and wellbore pressure have reached quasi-equilibrium, leak-off volume into formation should equal the shrinkage of fracture volume plus the expansion of wellbore fluid. We can modify the leak-off rate from Eq.(10) to account for wellbore fluid expansion as follows:

$$q_f = -\frac{A_f}{S_f} \frac{dP_f}{dt} - V_w \, c_w \frac{dP_f}{dt} = -\frac{A_f}{\frac{S_f A_f}{A_f + S_f V_w \, c_w}} \frac{dP_f}{dt} \tag{15}$$

Where $V_w$ is half the wellbore volume (only one wing of the fracture needs to be modeled) and $c_w$ is the compressibility of water. Using Eq.(15) with Eq.(10), we can define the fracture-wellbore system stiffness as:

$$S_s = \frac{S_f A_f}{A_f + S_f V_w \, c_w} \tag{16}$$

Under extreme circumstances, when wellbore storage is negligible ($V_w \, c_w \approx 0$), fracture stiffness dominates the system stiffness, and when $V_w \, c_w$ are large and $A_f$ is small, WBS dominates the system stiffness. If WBS has to be included in the model, one just need simply replace $S_f$ with $S_s$ in Eq.(12). The system stiffness reflects the overall compressibility of fracture and wellbore system.

## 2.2 Special Case 1: Constant Fracture Compliance and Carter's Leak-off

If we assume that a constant shut-in pressure (ISIP) is applied at the fracture surface, the solution of Eq.(6) to obtain the leak-off or the velocity $U_f$ across fracture surface into an intact, homogenous formation can be found as (Economides and Nolte 2000):

$$U_f = (\text{ISIP} - P_0) \sqrt{\frac{k\phi c_t}{\pi \mu_f \Delta t}} \tag{17}$$

If we take $\Delta t$ out of Eq.(17), then the R.H.S of Eq.(17) essentially becomes Carter's leak-off coefficient $C_L$, which is a constant parameter. The total leak-off rate $q_f$ from one wing of the fracture starting from shut-in is:

$$q_f = 2A_f(\text{ISIP} - P_0) \sqrt{\frac{k\phi c_t}{\pi \mu_f \Delta t}} \tag{18}$$

Combining Eq.(18) and Eq.(10), we have:

$$\frac{dP_f(\Delta t)}{d\Delta t} + 2S_f(\text{ISIP} - P_0) \sqrt{\frac{k\phi c_t}{\pi \mu_f \Delta t}} = 0 \tag{19}$$

Eq.(19) is a simple first order differential equation, with the initial condition of $P_f(\Delta t = 0) = \text{ISIP}$, the solution of Eq.(19)

leads to the pressure declining proportionally to the square root of time:

$$P_f(\Delta t) = \text{ISIP} - 4S_f(\text{ISIP} - P_0)\sqrt{\frac{k\phi c_t \Delta t}{\pi \mu_f}} \qquad (20)$$

Eq.(20) is the fundamental basis for the square root of time plot, which is equivalent to Notle's G-function model. From Eq.(1) and (20), we can infer that for normal-leak-off behavior with constant fracture stiffness and assuming Carter's leak-off, the pressure and its derivative will be straight lines on both G-function and the square root of time plots. However, a closer observation at Eq.(1) and Eq.(20), we can infer that the fracture pressure continues declining even when it reaches the initial reservoir pore pressure. This non-physical prediction indicates that the assumption of Carter's leak-off and constant stiffness made in the derivation prevent Notle's G-function model and the square root of time model from capturing the actual pressure response during fracture closure and late time pressure transient behavior.

## 2.3: Constant Fracture Compliance and Fracture Pressure Dependent Leak-off

When the leak-off rate is fracture pressure dependent, a closed analytic form of pressure decline is not available, but the solution can be obtained using a superposition time. A detailed derivation of the pressure decline solution is presented in **Appendix** . The fracture pressure at the $n^{\text{th}}$ time interval, $P_{f,n}$, can be calculated explicitly:

$$P_{f,n} = \text{ISIP} - 4S_f\sqrt{\frac{k\phi c_t}{\pi \mu_f}} \sum_{n=1}^{n-1}\sum_{j=1}^{n-1}(P_{f,j} - P_{f,j-1})\left(\sqrt{\Delta t_n - \Delta t_{j-1}} - \sqrt{\Delta t_{n-1} - \Delta t_{j-1}}\right) \qquad (21)$$

where $P_{f,0} = P_0$, $P_{f,1} = ISIP$ , $\Delta t_1 = 0$. It is important to note that in this article, the term "fracture pressure dependent leak-off" (FPDL) specifically refers to a leak-off rate that depends on the declining fluid pressure inside the fracture, not to how the leak-off rate changes with net stress in the rock surrounding the fracture, which is often denoted as pressure dependent leak-off or PDL (Barree et al. 2009). In some cases, excessive pressure drop can be observed during early-time of shut-in in low permeability reservoirs, but it is the result of high apparent ISIP, wellbore storage and additional friction away from the wellbore, and not caused by PDL, which is unlikely to happen in unconventional reservoirs.

## 2.4 Variable Fracture Compliance

### 2.4.1 The Cause of Variable Fracture Compliance

There are basically two main causes that lead the continuously changing of fracture compliance during closure. The first is stress contrast across different layers that the fracture has penetrated into. In this case, fracture will close first in the zones where the minimum in-situ stress is highest, which alters the overall fracture stiffness during the closure process. The second cause of variable fracture compliance is fracture surface asperities and roughness, where the fracture closes on asperities progressively from its edges to the center, and the overall fracture stiffness is determined by both the closed portion and open portion of fracture during closure.

As pressure declines inside the fracture after shut-in, the fracture will gradually close and the fracture aperture will approach the scale of the surface roughness. If the fracture faces are perfectly parallel and smooth, they will come into contact all at once when the fluid pressure inside the fracture declines to the far field stress, and the fracture is then mechanically and hydraulically closed. However, there is abundant evidence to suggest that fractures retain their conductivity after the walls have come into contact (mechanical closure). Fractures retain a finite aperture after mechanical closure due to a mismatch of asperities on the fracture walls. van Dam et al. (2000) presented scaled laboratory experiments on hydraulic fracture closure behavior. They observed up to a 15% residual aperture (compared to the maximum aperture during fracture propagation) long after shut-in. Fredd et al. (2000) demonstrated fracture surface asperities can provide residual fracture width and sufficient conductivity in the absence of proppants. Using sandstone cores from the East Texas Cotton Valley formation, sheared fracture surface asperities that had an average height of about 2.286 mm were observed. Warpinski et al. (1993) reported hydraulic fracture surface asperities of about 1.016 mm and 4.064 mm for nearly homogeneous sandstones and sandstones with coal and clay-rich bedding planes, respectively. Sakaguchi et al. (2008) created tensile fractures in large rock blocks and measured the asperity height and distribution. Their work shows that the fracture surfaces can be assumed to be a fractal object and most of the asperities fall within 1 to 2 mm in height. Wells and Davatzes (2015) conducted topographic measurement on dilated fractures from core samples and found the asperity height ranges from hundreds to thousands of micrometers. Bhide et al. (2014) created X-ray microtomographic images from shear induced fractures and the roughness values obtained varied from 1.8 to 1.95 mm along the length of the rock samples. Zou et al. (2015) conducted experiments on 20 fractured shale samples and found the average asperity height to be 1.88 mm. An experimental study (Zhang et al. 2014) on Barnett Shale samples reveals that the surface topography of the displaced fracture can be altered because of rock failure, and the fracture surface exhibited parallel strips of crushed asperities. Field measurements (Warpinski et al. 2002) using a down-hole tiltmeter array indicated that the fracture closure process is a smooth, continuous one which often leaves 20%-30% residual fracture width, regardless of whether the injection fluid is water, linear-gel or cross-linked-gel.

The microscopic measurement and modeling of surface roughness and mechanical properties of asperities can often be up-scaled to macroscopic contact laws that relate fracture width and the associated contact stress. Willis-Richards et al. (1996) proposed contact law to relate fracture width and the net closure stress for fractured rocks, based on the work of Barton et al. (1985):

$$\sigma_c = \frac{\sigma_{ref}}{9}\left(\frac{w_0}{w_f} - 1\right) \text{ for } w_f \leq w_0 \tag{22}$$

where $w_f$ is the fracture aperture and, $w_0$ is the contact width, which represents the fracture aperture when the contact normal stress is equal to zero, $\sigma_c$ is the contact normal stress on the fracture, and $\sigma_{ref}$ is a contact reference stress, which denotes the effective normal stress at which the aperture is reduced by 90%. it should be emphasized that the contact width $w_0$ is determined by the tallest asperities, and the strength, spatial and height distribution of asperities are reflected by the contact reference stress $\sigma_{ref}$ (e.g., if the tallest asperities on two fracture samples are the same, then they should have the same $w_0$, but the one with a higher median asperity height or Young's modulus will have higher value of $\sigma_{ref}$, provided other properties are the same).

## 2.4.2 Determine Pressure Dependent Fracture Compliance

With known fracture geometry, rock properties and surface roughness (represented by contact parameters $w_0$ and $\sigma_{ref}$), the question now is how to estimate fracture stiffness (or compliance) as a function of pressure, if it continuously changes during fracture closure. Once the pressure dependent fracture stiffness is obtained, it can be substituted into Eq.(12) and the whole system of equations can be solved readily. Wang and Sharma (2017) presented an integral transform method and general algorithms to model the dynamic behavior of hydraulic fracture closure on rough fracture surfaces and asperities, using linear elastic solutions that coupled with contact law for three different fracture models (PKN, KGD and radial fracture geometry). Given the fracture geometry, rock properties, contact parameters and minimum principal stress, their approach can predict the evolution of fracture aperture profile, total fracture volume and fracture stiffness as fracturing pressure declines. Wang et al. (2017) presented an improved model for fracture closure based on superposition principles. Their model can simulate large scale fracture closure behavior with layer stress contrast in an efficient manner.

Detailed modeling of non-local fracture closure on asperities and rough surfaces has already been discussed extensively (Wang and Sharma 2017; Wang et al. 2017), hence will not be discuss further. Here, we'll examine an example of a fracture that closes on asperities and how the fracture stiffness evolves during closure. Assuming a Young's modulus of 20 GPa, Poisson's ratio of 0.25, $w_0$ of 2 mm, $\sigma_{ref}$ of 5 MPa for a PKN fracture geometry with 10 m fracture height, 35 MPa minimum in-situ stress, the evolution of the fracture width profile and contact stress distribution can be determined, as the fluid pressure inside the fracture gradually declines. The results are shown in **Fig.3** and **Fig.4**. To demonstrate the impact of fracture roughness and surface asperities on fracture closure behavior, the case without surface asperities (fracture surface is completely planar and smooth) is also included. The result shows that at relative high fracturing fluid pressure, the fracture asperities have negligible impact on fracture width distribution, and the contact stress is always concentrated at the tip of the fracture, where the contact stress is much higher than in the middle of the fracture. We can also see that the fracture surfaces do not contact each other like parallel plates. In fact, the fracture closes on rough surfaces starting from the tip, and closes progressively all the way from the edges to the center of fracture. As fluid pressure continues to decline, more and more of the fracture surfaces come into contact and these changes the subsequent fracture closure behavior. At lower fluid pressures, contact stresses start to counter-balance the in-situ stress and the fracture becomes stiffer and less compliant. If the fracture faces were perfectly parallel and smooth, the fracture width would have collapsed to zero when the fluid pressure dropped to 35 MPa. The moment when all fracture surfaces come into contact on asperities and the contact stress becomes non-zero on the entire fracture surface, the fracture is mechanically closed and this mechanical closure stress is higher than the minimum in-situ stress.

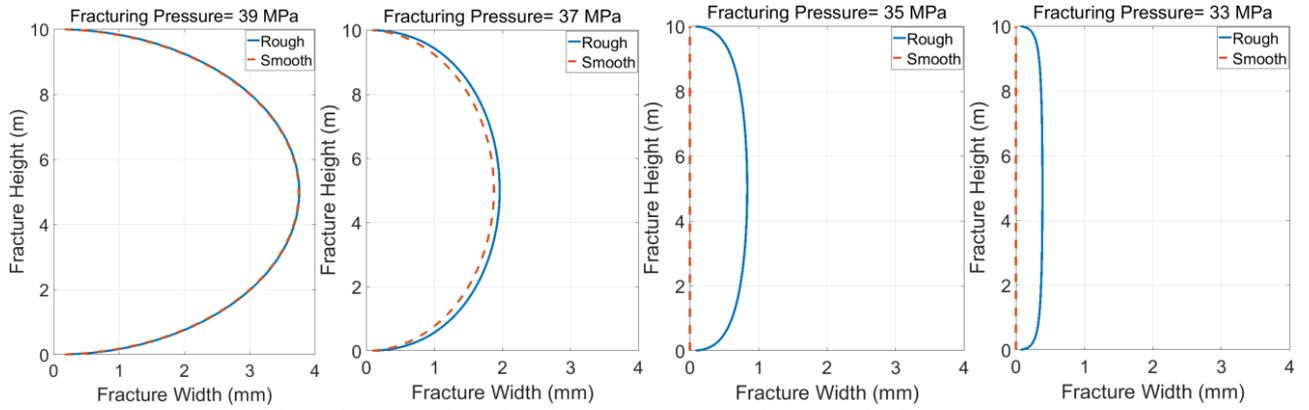

**Fig.3 Fracture width evolution with and without asperities at different fluid pressure for a PKN geometry**

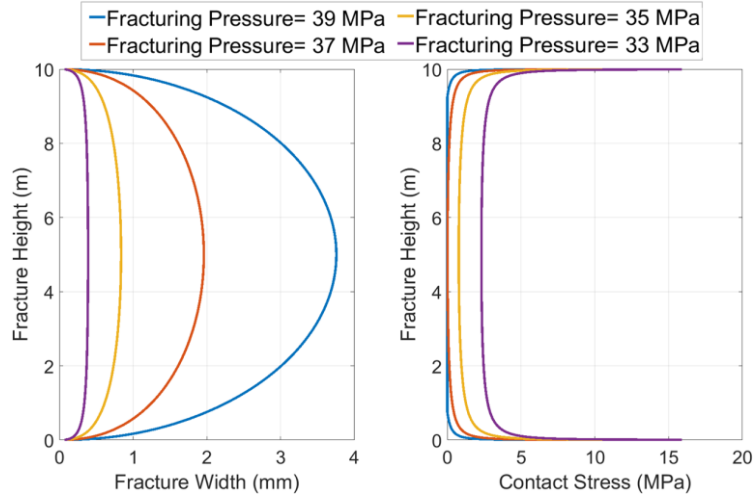

**Fig.4 Fracture width and the corresponding contact stress distribution at different fluid pressure for a PKN geometry**

The simulated fracture volume evolution as a function of fluid pressure is shown in **Fig.5**. As can be seen, when the fluid pressure inside the fracture is relatively high, the fracture volume declines linearly with pressure (indicates roughly constant fracture compliance/stiffness). However, as the pressure declines to a certain level (2 MPa higher than the input minimum in-situ stress of 35 MPa), the fracture volume and pressure departs from a linear relationship. The traditional method of estimating fracture stiffness using Table 1 can only be used when the fracture pressure is still relatively high and the asperities are only in contact at the edges of the fracture. However, as pressure declines further and more and more fracture surface area comes into contact, the fracture stiffness becomes pressure dependent:

$$S_f = \frac{A_f dP_f}{dV_f} \tag{23}$$

The relationship between $P_f$ and $V_f$ from the output of non-local fracture closure model can be used to calculate the pressure dependent fracture stiffness and the result is also shown in Fig.5. This pressure dependent fracture stiffness can be then incorporated into Eq.(12) to make $S_f$ pressure dependent. The "tangent line method" (our using the end of 3/2 slope on log-log plot) for picking up minimum in-situ stress is based on the underlying assumptions of G-function model that assumes perfectly smooth fracture surfaces, and the stiffness remains constant until fluid pressure drops to minimum in-situ stress. This leads to a sudden increase in fracture stiffness (or fracture compliance abruptly drop to zero) when all fracture surfaces get into contact simultaneously. Fracture stiffness only remains constant at high fracture pressures. It starts to increase long before the fracturing pressure declines to the minimum in-situ stress (35 MPa in this case).

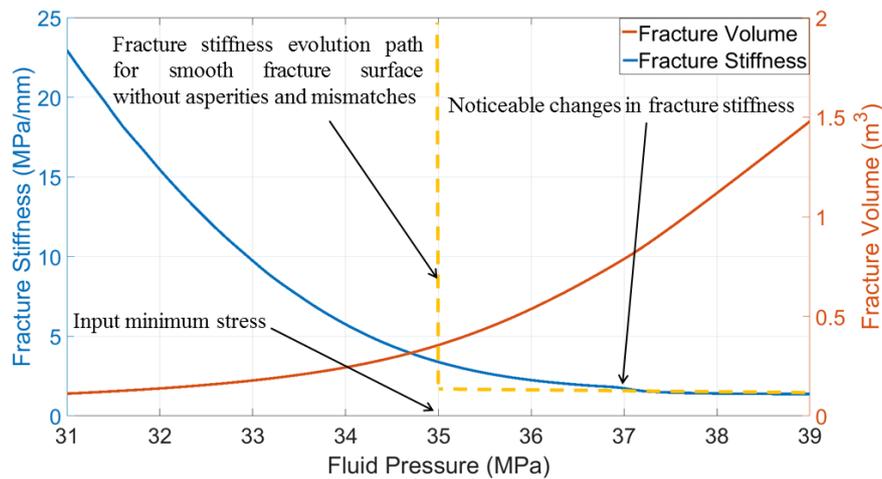

**Fig.5 Fracture volume and fracture stiffness evolution as fluid pressure declines for a PKN geometry**

There is compelling field evidence to indicate that this occurs in the field. **Fig.6** shows the normalized tiltmeter response plotted against wellbore pressure during the shut-in period of 2B well at three different stations from the GRI/DOE M-site. Each tiltmeter response is normalized by dividing by the maximum tilt (displacement, also an indication of fracture width) measured at that instrument during the test. The tiltmeter demonstrates that soon after shut-in, the measured displacement declines linearly with pressure. If the fracture surface area remains constant during shut-in, then the fracture volume is proportional to the average displacement and should also decline linearly with the pressure within this period. After the wellbore pressure declines to a certain level, the measured displacement vs pressure departs from a linear relationship. This field measurement is consistent with the general trend shown in Fig.5. As pressure continues declining, more and more fracture surface area comes into contact with the rough fracture surfaces and asperities, and the fracture stiffness increases gradually. Even though these data were measured from the end of injection to weeks of shut-in, we can still observe the existence of a residual fracture width that is supported by surface asperities and mismatches even after the fracturing fluid pressure drops below the minimum in-situ stress. So considering the nature of mineralogical heterogeneity and the existence of thin laminated rock layers, the formation of asperities and tortuous fracture walls are inevitable during hydraulic fracture propagation.

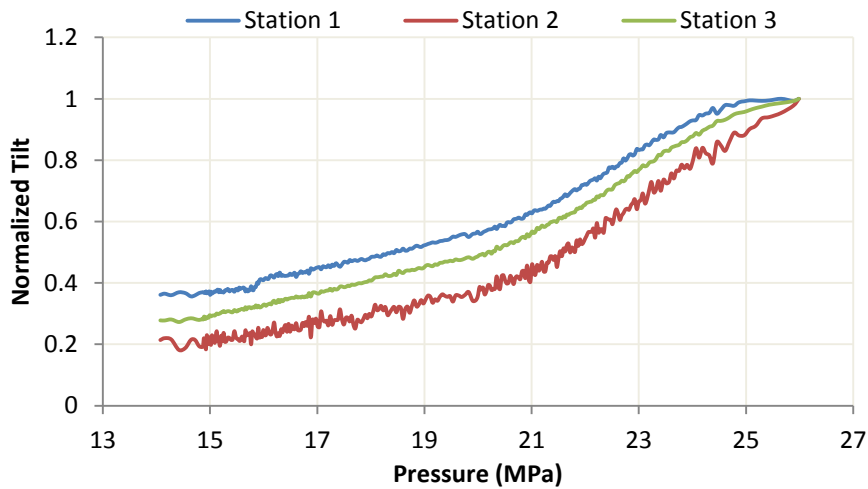

**Fig.6 Normalized tiltmeter data from shut-in of 2B from the GRI/DOE M-site. Data Courtesy of Norm Warpinski**

## 3. Simulation Results and Analysis

The pressure decline behavior during DFITs is governed by Eq.(6) (a PDE) with Eq.(12) (an ODE) acting as the boundary condition. To solve this system of PDE with an ODE boundary condition, the method of lines (MOL) is used, by replacing the spatial derivative in the PDE with algebraic approximations. The PDE can then be transformed into a system of ODEs, which can be solved simultaneously and efficiently by well-established numerical methods (Schiesser and Griffiths 2009). Since the MOL essentially replaces the problem PDEs with systems of approximating ODEs, the addition of other ODEs is easily accomplished. In this section, our global DFIT model is used to investigate how different fracture geometry, surface roughness, reservoir properties and stress contrast impact fracturing pressure response and our interpretation.

### 3.1 Normal leak-off behavior

Normal leak-off behavior is a special case of our model that satisfies all the assumptions of Nolte's G-function model, and it can be clearly identified with straight lines of pressure and its derivatives on G-function and the square root of time plots. All previous BCA models assume that these straight lines indicate normal leak-off behavior and hence, leak-off coefficient or formation permeability can be determined based on this portion of the data. However, two key assumptions in all BCA are Carter's leak-off and constant fracture compliance. To examine the impact of incorrectly assuming Carter's leak-off for the duration of the DFIT, we extend the concept of "normal leak-off behavior" to account for different fracture boundary conditions, while other assumptions remain the same (i.e., constant fracture compliance and constant surface area during closure). The simulation results from three different leak-off models are compared using input parameters from **Table 2**. **Figures 7** and **8** show the simulated pressure and its derivatives on G-function and the square root of time plots for different formation permeability. As expected, Carter's leak-off (using Eq.(1) or Eq.(20), they are interchangeable) always produce straight lines on diagnostic plots. But when pressure dependent leak-off at the fracturing surface is considered (using Eq.(21) and our fully coupled simulation), we no longer obtain straight lines. This is because as fracturing pressure declines, leak-off rate departs from the linear relationship with $1/\sqrt{\Delta t}$, so the assumption of Carter's leak-off is violated. Carter's leak-off model overestimates the leak-off rate and pressure drop by assuming a constant ISIP at the fracturing surface and leads to larger errors as shut-in time gets longer.

| Fracture type | PKN |
|---|---|
| Fracture height | 10 m |
| Fracture length | 50 m |
| Pumping time | 5 min |
| ISIP | 40 MPa |
| Minimum in-situ stress | 35 MPa |
| Initial pressure | 20 MPa |
| Young's modulus | 20 GPa |
| Total compressibility | 1.9e-3 MPa$^{-1}$ |
| Viscosity | 1 cP |
| Poisson's Ratio | 0.25 |
| Initial porosity | 0.03 |

**Table 2-Input parameters for DFIT simulation with constant fracture compliance**

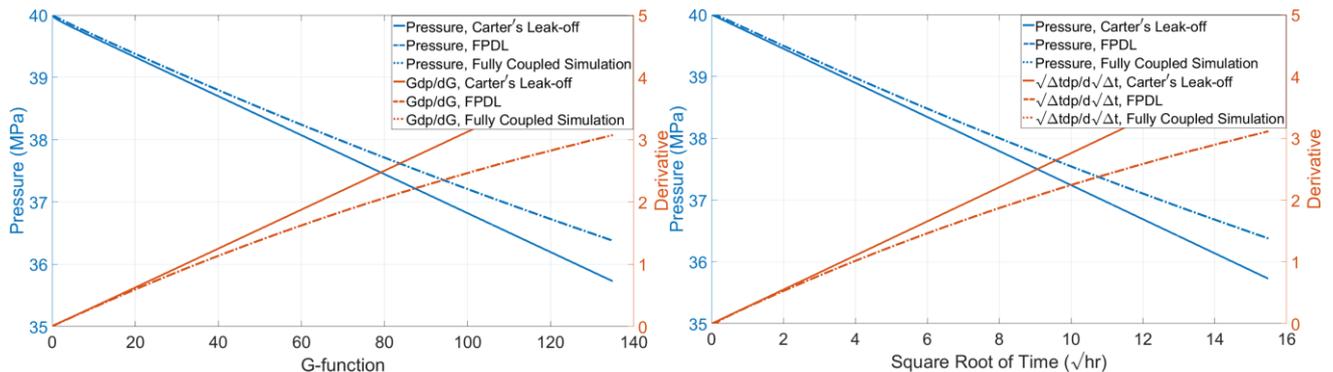

**Fig.7 Comparison of "normal leak-off" behavior for different models, k=0.0001 md**

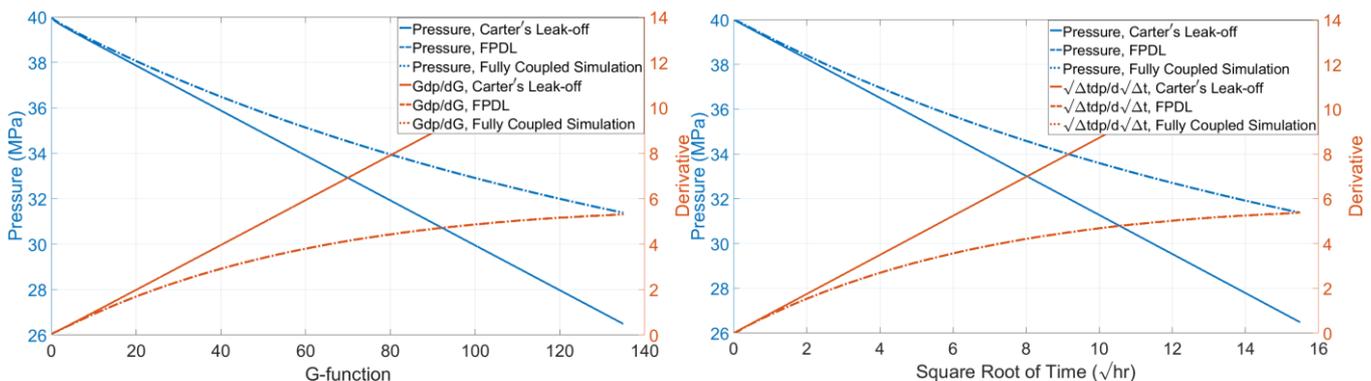

**Fig.8 Comparison of "normal leak-off" behavior for different models, k=0.001 md**

The most important observation here is that the so-called "normal leak-off behavior" with straight lines on G-function and the square root of time plots is not "normal" at all. For example, the closure time for typical DFIT in a formation with 0.0001 md permeability is around 2~4 days. And in the above case shown in Fig.7, the fracturing pressure and its derivative depart from the straight line after 4 hours of shut-in, even with constant fracture compliance/stiffness, surface area and reservoir properties. Many field data sets do show "normal leak-off behavior" with straight lines that extend all the way to the point of downward deflection. This occurs if some mechanisms that accelerate pressure drop (e.g., the gradual increase of fracture stiffness) happen to offset the effects of declining pressure at the fracture surface, so the pressure data fits the straight lines predicted by Carter's leak-off model coincidentally. So any DFIT analysis that is based on the assumption of Carter's leak-off and straight lines on G-function and the square root of time plots should be re-examined.

### 3.2 Variable Fracture Compliance on Pressure Decline and In-Situ Stress Estimation

As discussed earlier, G-function and previous BCA models assume Carter's leak-off behavior as if the pressure in the fracture is a constant value (i.e., ISIP). In addition, these models also assume fracture compliance/stiffness is constant and only changes abruptly at the moment of fracture closure, when fracturing pressure drops to the minimum in-situ stress. The failure to account for the coupled mechanisms of pressure dependent leak-off at the fracture surface and the gradual increase of fracture stiffness during closure explains why these models are only useful on a small portion of the before closure data, and are not able to simulate the fracturing response for the entire duration of the DFIT test. These simplifying assumptions also make their interpretation of DFIT data and estimation of minimum in-situ stress inaccurate. In this section, our global DFIT model is used to investigate how each factor impacts the pressure decline trend and our estimation of minimum in-situ stress and fracture compliance.

#### 3.2.1 PKN Fracture Geometry

We first examine a PKN geometry using parameters listed in Table 2 and assuming $w_0$ is 2 mm, $\sigma_{ref}$ is 5 MPa and reservoir permeability is 0.0005 md as the Base Case. **Fig.9** shows the contact stress at different fracture width. As expected, when the fracture width is larger than the contact width $w_0$, the contact stress is zero. However, when the fracture width is smaller than the contact width, the contact stress and fracture width follows a hyperbolic relationship. **Fig.10** shows the corresponding fracture stiffness evolution for different $w_0$ with the given fracture geometry and rock properties. The results indicate that as the fracture width decreases, the rough fracture walls will come into contact sooner if the contact width is larger, so the noticeable changes of fracture stiffness occur earlier when the contact width is larger. We can also observe that when the contact width is larger, the increase in fracture stiffness is more gradual and smooth. In the extreme case of $w_0$ close to zero, as would be the case for relative flat fracture surfaces, the fracture stiffness will increase abruptly as soon as the fluid pressure drops to the minimum in-situ stress (35 MPa in this case). This pressure dependent fracture stiffness obtained from non-local fracture modeling (Wang and Sharma 2017) can be directly put into the boundary ODE of Eq.(12) to predict fracture pressure decline for certain reservoir properties.

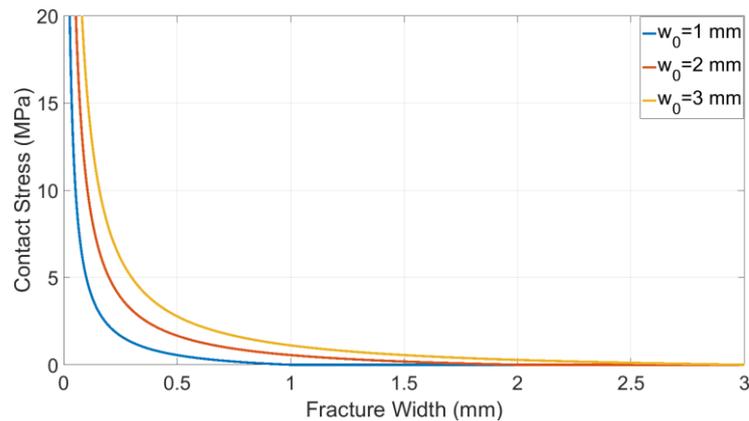

**Fig.9 The relationship between contact stress and fracture width for different $w_0$**

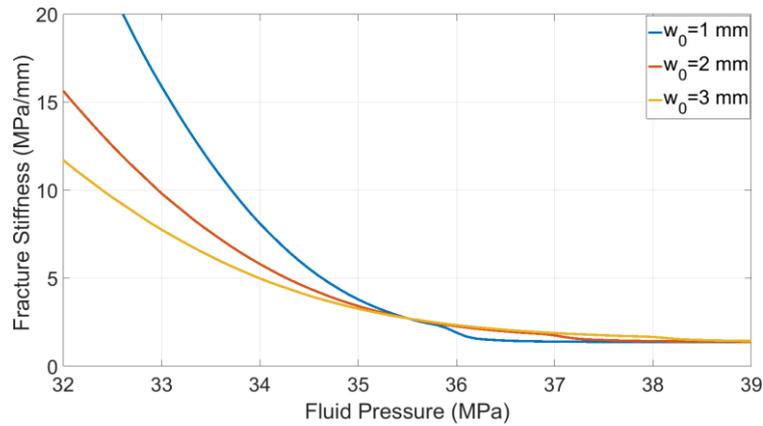

**Fig.10 Fracture stiffness evolution for different $w_0$ with a PKN geometry**

**Fig.11** shows the fracturing pressure and its derivatives for different $w_0$ on G-function and square root of time plots. Relate with Fig.9 and Fig.10, we can see that the contact width impacts pressure decline response significantly, because it alters the evolution of fracture stiffness. Large contact width leads to a smooth pressure decline trend while small contact width leads to steep changes in pressure decline rate and pressure derivatives. We can also infer that if the contact width is close to zero and all fracture walls come into contact simultaneously, then a sudden change in pressure decline rate and the pressure derivative spikes on both G-function and square root of time plots are inevitable, which is unrealistic and never observed in field cases. So the conventional assumption that a fracture closes on flat, smooth fracture surfaces where $w_0 = 0$ does not reflect reality. In addition, different contact widths lead to different derivative slopes, so any BCA models that attempt to match this portion of data and estimate reservoir properties without considering the variable fracture compliance is questionable. Fig.11 also shows G-function and square root of time plots give the same quantitative information only slightly different in scales. More importantly, the estimated minimum in-situ stress from pressure derivatives using the "convention method" and "compliance method" give inconsistent results for different contact widths, with the same input parameters of fracture surface area, minimum in-situ stress and reservoir properties. A detailed comparison and a new approach to estimate minimum in-situ stress will be discussed later.

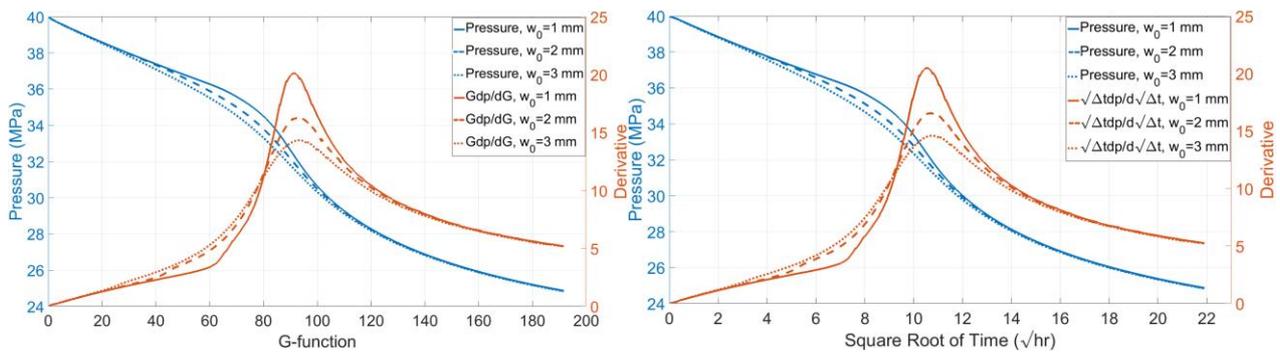

**Fig.11 Pressure decline response for different $w_0$ with a PKN geometry**

Next, we examine how contact reference stress affects the pressure decline response. **Fig.12** and **Fig.13** show the relationship between contact stress and fracture width for different contact reference stress and the corresponding fracture stiffness evolution at different fracturing pressure. For the same contact width, the higher the contact reference stress, the more rapid the increase of contact stress as the fracture width shrinks. Physically, the contact reference stress represents how hard and strong the fracture surface asperities are. The lower the contact reference stress, the more gradual the change in fracture stiffness as pressure declines. Even though the contact reference stress does not have much impact on the pressure at which the fracture stiffness starts to changes noticeably, it does impact the fracture stiffness evolution, as shown in Fig.13. **Fig.14** shows the fracturing pressure and its derivatives for different contact reference stress on G-function and the square root of time plots. It can be observed that the contact reference stress has negligible influence on early time pressure decline when fracturing pressure is still high. This is because only the fracture edges start to come into contact during this period and the fracture stiffness remains roughly constant. However, after this period as the fracture pressure declines further, more and more fracture surfaces come into contact and the contact reference stress begins to affect the pressure decline trend and the peak value of the pressure derivatives.

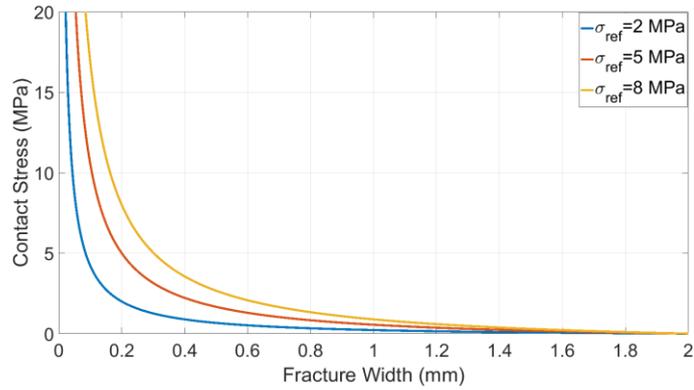

**Fig.12 The relationship between contact stress and fracture width for different $\sigma_{ref}$**

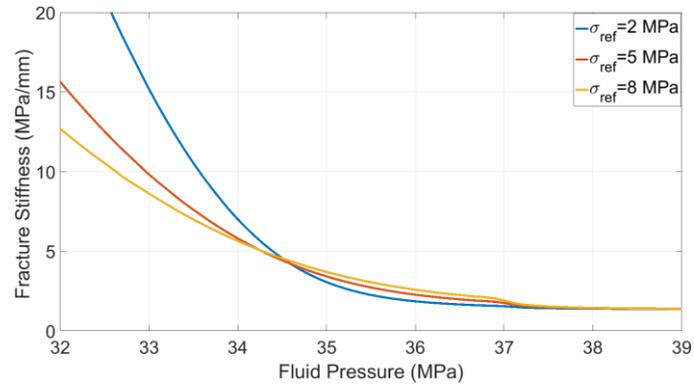

**Fig.13 Fracture stiffness evolution for different $\sigma_{ref}$ with a PKN geometry**

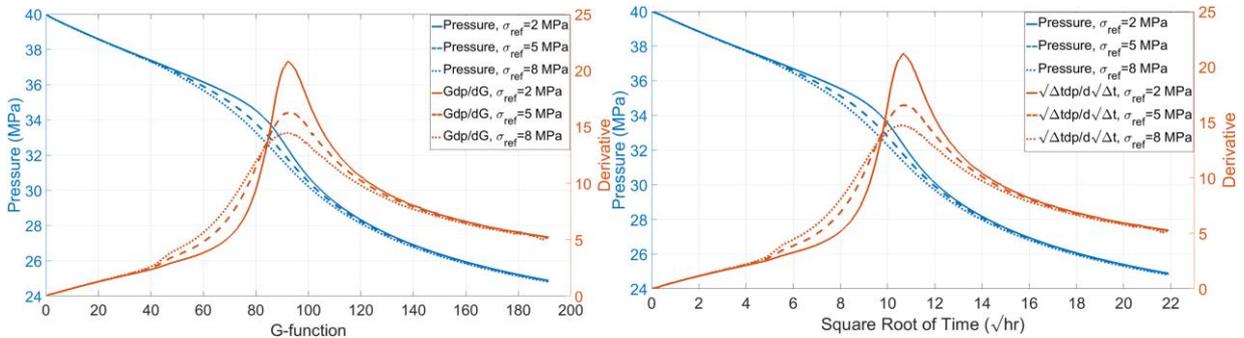

**Fig.14 Pressure decline response for different $\sigma_{ref}$ with a PKN geometry**

Besides the contact parameters, fracture geometry also affects fracture stiffness and its evolution during closure. **Fig.15** shows fracture stiffness evolution for different fracture height while all the other parameters remain the same as the Base Case. Similar to what Table 1 implies, for a PKN fracture geometry, the smaller the fracture height, the higher the initial fracture stiffness. We can also observe that smaller fracture height leads to noticeable changes in fracture stiffness at higher fracturing pressure. This is because smaller fracture height results in smaller fracture width at the same net pressure, so the more fracture surfaces can come into contact at a higher fracture pressure. When fracture height is 5m, the effective fracture stiffness is already affected by surface contact when fluid pressure is 39 MPa.

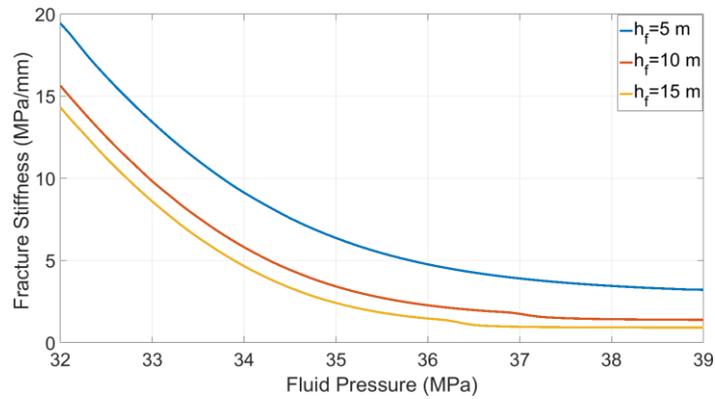

**Fig.15 Fracture stiffness evolution for different fracture height with a PKN geometry**

**Fig.16** shows the fracturing pressure and its derivatives for different fracture height on G-function and the square root of time plots. The results indicate that fracture height impacts the pressure decline trend significantly. Larger fracture height leads to later occurrence of the peak of the pressure derivative and this also increases the peak value of the pressure derivatives.

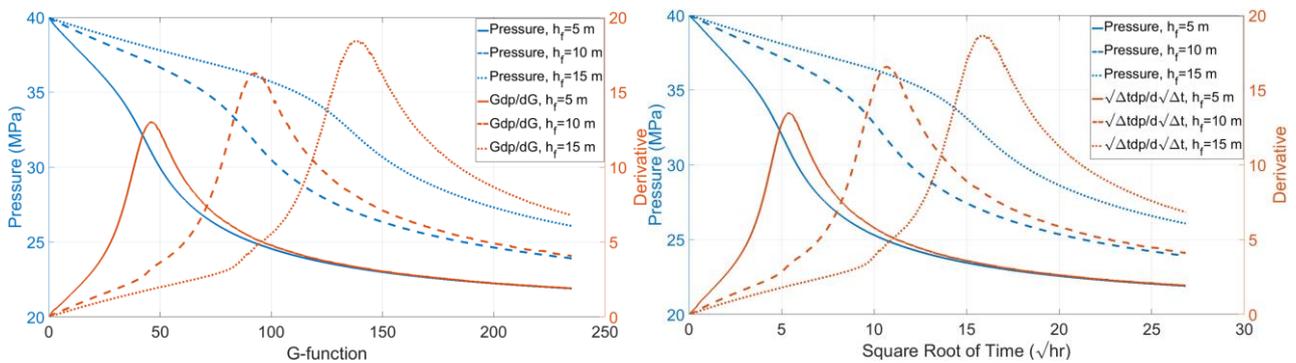

**Fig.16 Pressure decline response for different fracture height with a PKN geometry**

**Fig.17** shows the fracturing pressure and its derivatives for different reservoir permeability on G-function and the square root of time plots. As expected, the pressure declines more rapidly when the reservoir permeability is large and the decline rate slows down as the difference between fracturing pressure and initial reservoir pressure becomes smaller and smaller.

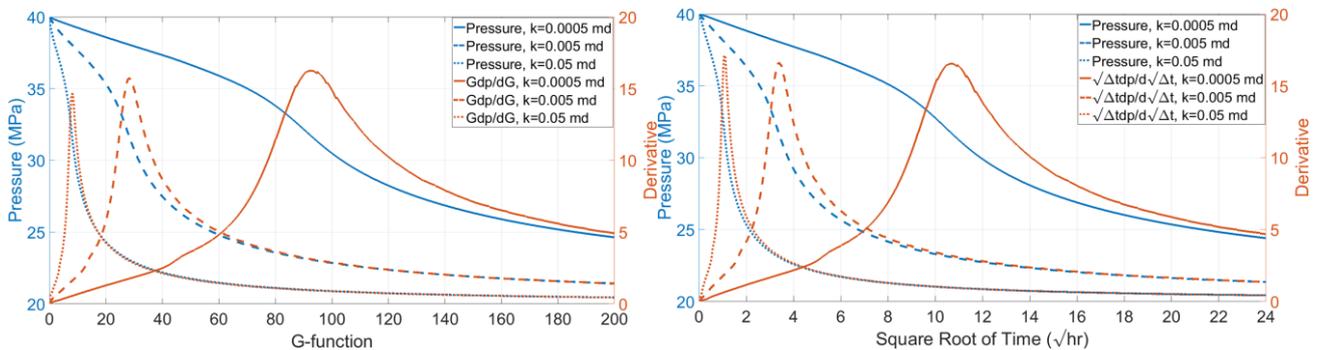

**Fig.17 Pressure decline response for different reservoir permeability with a PKN geometry**

Next, we examine the impact of wellbore storage on the fully coupled pressure decline trend. The water compressibility is assumed to be 4.35e-4 MPa$^{-1}$. **Fig.18** shows the fracture-wellbore system stiffness evolution for different wellbore volume. As can be seen, when fracturing pressure is high, fracture stiffness dominates the system stiffness. However, as fracturing pressure continues declining, the fracture become less and less compressible and the role of wellbore storage becomes apparent. In general, the larger the wellbore volume, the more gradual and slower the increase in system stiffness will be.

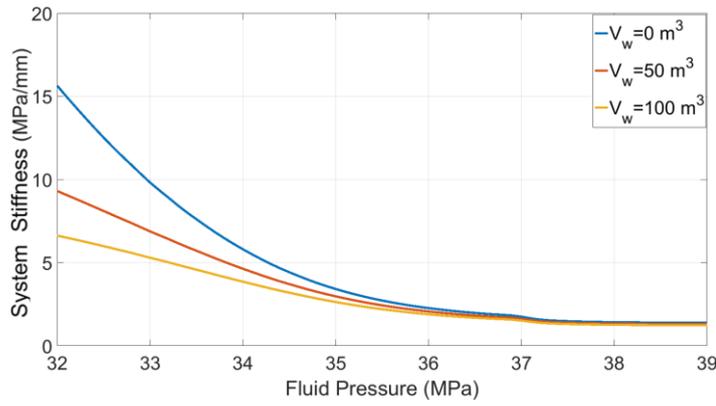

**Fig.18 Fracture-wellbore system stiffness evolution for different wellbore volume with a PKN geometry**

**Fig.19** shows the corresponding fracturing pressure and its derivatives for different wellbore volumes on G-function and square root of time plots. It can be observed that larger wellbore volume leads to more gradual pressure decline trends. A larger wellbore volume also delays the occurrence of fracture closure and lowers the peak of the pressure derivative curve. It can be seen that WBS has a small impact during early time of shut-in when the system stiffness is still dominated by fracture stiffness, however, as more and more of the fracture surface comes into contact and the fracture becomes stiffer, WBS effects become apparent, and the after-flow of fluid from wellbore to fracture long after shut-in decelerates the pressure decline rate and extends the tail of the pressure derivative after its reaches the peak. So the WBS effect on pressure decline response must be accounted for when interpreting and analyzing DFIT data.

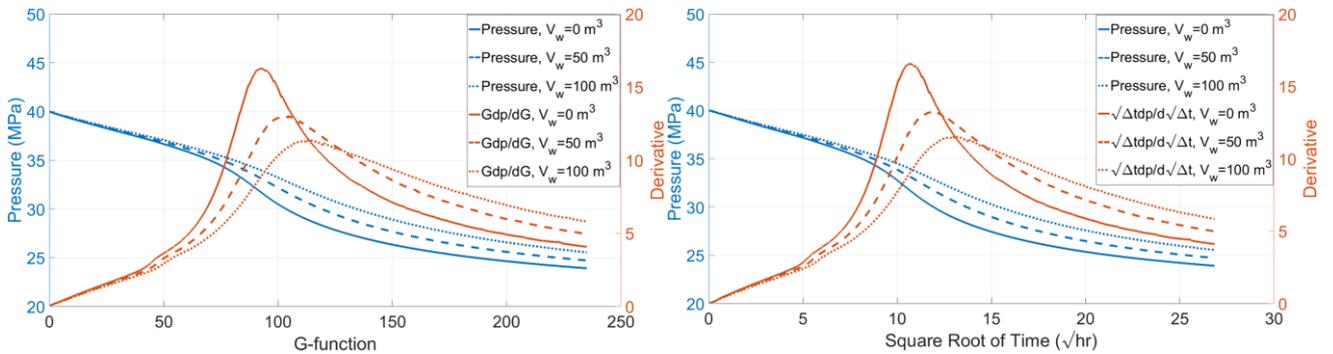

**Fig.19 Pressure decline response for different wellbore volume with a PKN geometry**

**Fig.20** shows the fracturing pressure and its derivatives for different initial reservoir pressure on G-function and square root of time plots. It can be observed when the initial reservoir pressure is low; the pressure declines more rapidly. This is also reflected by Eq.(18), a lower reservoir pressure leads to a higher leak-off rate. However, when the reservoir pressure is over pressurized with high initial pore pressure, the pressure decline trend resembles a "normal-leak-off behavior". In this case ($P_0$=28MPa), one can still notice that there exist a subtle "bump" in the pressure derivatives that indicates an increase in fracture stiffness as reflected in Fig.10. For small values of $\sigma_{ref}$, this increase can be too gradual to be noticeable on G-function and square root of time plots. This example demonstrates that long closure time not only can be attributed to low formation permeability, but also can be the result of a small pressure difference between ISIP and initial pore pressure.

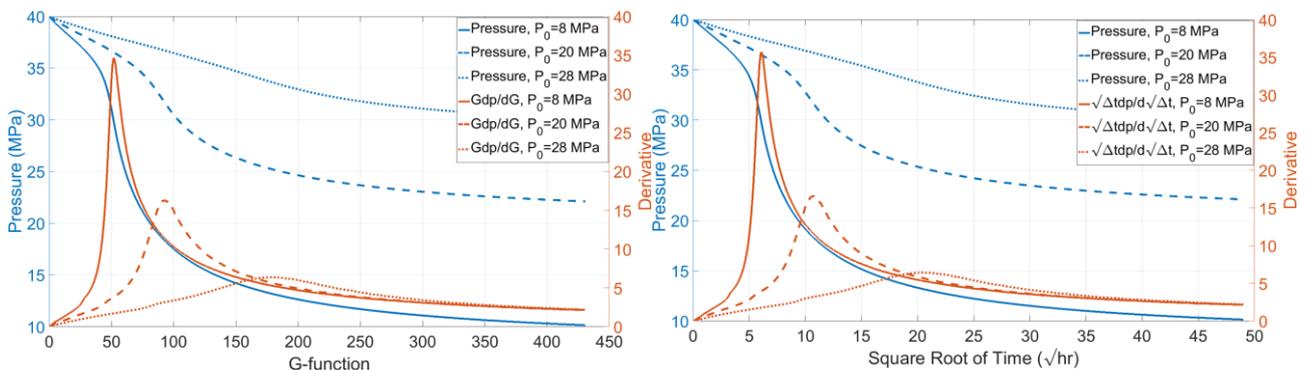

**Fig.20 Pressure decline response for different initial reservoir pressure with a PKN geometry**

In all the above synthetic simulations, the input minimum in-situ stress remains 35 MPa while other parameters are modified from the Base Case. The pressure decline trend varies substantially for different fracture geometry, contact parameters, reservoir permeability, wellbore volume and reservoir pore pressure. Any interpretation of the pressure data to determine the minimum in-situ stress must properly account for all these parameters. Choosing minimum in-situ stress where the pressure or its derivative departs from a straight line (Nolte 1979 and1986) is not reliable, because it ignores the changes in fracture compliance/stiffness during closure and the pressure dependent leak-off at the fracture surface. In addition, pressure does not decline linearly with G-function or $\sqrt{\Delta t}$ in most cases. The most widely used "tangent line method" to pick up minimum in-situ stress using a tangential line with pressure derivatives (Barree et al., 2009; Barree et al., 2014) always underestimates minimum in-situ stress, especially in under-pressure reservoirs. Even though this method has been the common practice in the industry, it is an emprical method and is subject to the assumptions discussed earlier. The "compliance method" identifies closure pressure as the point at which fracture stiffness starts to change (McClure et al. 2016). As we have discussed earlier, the fracture closes progressively from the edges to the center, As a result the fracture stiffness begins to increase long before the fracture pressure drops to the minimum in-situ stress. A comparison of different methods of estimating minimum in-situ stress (closure pressure is interpreted as minimum in-situ stress) for all previous synthetic cases is presented in **Table 3**.

| | Scenarios | Tangent Line Method | | Compliance Method | | Variable Compliance Method | |
|---|---|---|---|---|---|---|---|
| | | Estimated $\sigma_{hmin}$ (MPa) | Absolute Error (MPa) | Estimated $\sigma_{hmin}$ (MPa) | Absolute Error (MPa) | Estimated $\sigma_{hmin}$ (MPa) | Absolute Error (MPa) |
| Modified Base Case | Base Case | 32.3 | -2.7 | 37.3 | 2.3 | 35.3 | 0.3 |
| | $w_0$=1 mm | 32.7 | -2.3 | 36.7 | 1.7 | 35.1 | 0.1 |
| | $w_0$=3 mm | 31.7 | -3.3 | 38.2 | 3.2 | 35.0 | 0 |
| | $\sigma_{ref}$=2 MPa | 32.4 | -2.6 | 36.9 | 1.9 | 35.2 | 0.2 |
| | $\sigma_{ref}$=8 MPa | 32.2 | -2.8 | 37.3 | 2.3 | 35.1 | 0.1 |
| | $h_f$=5 m | 31.3 | -3.7 | 39.3 | 4.3 | 35.4 | 0.4 |
| | $h_f$=15 m | 32.5 | -2.5 | 36.8 | 1.8 | 35.1 | 0.1 |
| | k=0.005 md | 32 | -3 | 37.4 | 2.4 | 35.2 | 0.2 |
| | k=0.05 md | 31.8 | -3.2 | 37.5 | 2.5 | 35.3 | 0.3 |
| | $V_w$=50 $m^3$ | 32.7 | -2.3 | 37.1 | 2.1 | 35.3 | 0.3 |
| | $V_w$=100 $m^3$ | 32.8 | -2.2 | 37.1 | 2.1 | 35.2 | 0.2 |
| | $P_0$=8 MPa | 30 | -5 | 37.4 | 2.4 | 35.3 | 0.3 |
| | $P_0$=28 MPa | 34.3 | -0.7 | NA | NA | NA | NA |

**Table 3-Comparison of different methods of estimating minimum in-situ stress for PKN fracture geometry**

Examining Eq.( 6)~Eq.(16) closely, it is evident that the pressure decline is governed by a linear PDE system in the formation that is coupled with an ODE at the fracture surface. The minimum in-situ stress is just one of many implicit factors that shape the evolution of the fracture pressure. Previous DFIT models simplified this PDE-ODE system assuming Carter's leak-off and constant fracture stiffness during fracture closure, so that a closed form of an analytic solution could be obtained. Since our model shows minimum in-situ stress implicitly coupled with pressure, it is not possible to analytically estimate the minimum in-situ stress from a single diagnostic plot. Based on the analysis of our synthetic cases, it is clear that the "tangent line method" underestimates minimum in-situ stress and the "compliance method" overestimates closure in all cases if the closure pressure is interpreted as the minimum in-situ stress.

Our analysis shows that the answer lies between these two limiting cases, regardless of parameters used in simulation. To keep the analysis pratical and not resorting to a complex inverse problem, we propose a "variable compliance method", where the minimum in-situ stress can be obtained by averaging the dimensionless G-time or the square root of time of these two methods and extrapolating back to the pressure curve that corresponds to the averaged G-time or the square root of time. An exception is when pressure derivative plot resembles the so-called "normal leak-off behavior", the changes in fracture fracture/compliance may not be detectable on diagnostic plots and the "tangent line method" gives an estimation of minimum in-situ stress within an acceptable error.

The estimated minimum in-situ stress using "variable compliance method" is also included in Table 3. When we compare the estimated minimum in-situ stress obtained from the different methods with the input minimum in-situ stress, Table 3 shows that the "variable compliance method" gives significantly lower errors. The maximum error does not exceed a few 100 KPa, while the "tangent line method" yield errors that are an order of magnitude larger.

Even though we have focused on the PKN fracture geometry in this section, these conclusions are also applicable to a KGD fracture geometry. We can obtain the results for a KGD geometry if we replace fracture height with fracture length in calculating the evolution of fracture stiffness, as discussed in Wang and Sharma (2017).

### 3.2.2 PKN Fracture Geometry with Multiple Layers

In previous section, we analyzed the pressure decline response of a DFIT in a single layer; however, it is possible that a fracture can penetrate into adjacent layers where the minimum in-situ stresses are different. **Fig.21** illustrates examples where the fracture spans across multiple layers. In a single layer case, the fracture gradually closes from edges to center on rough fracture walls, while in multilayer cases, the fracture can close first within the interval where the minimum in-situ stress is the largest.

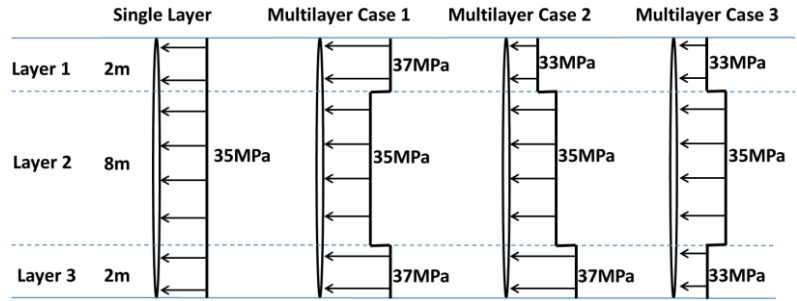

**Fig.21 Illustration of multilayer cases with different minimum in-situ stresses**

Assuming the input parameters are the same as the base case in the previous section (provided in Table 2), and fracture geometry and minimum in-situ stresses are illustrated in Fig.21, the fracture stiffness evolution can be calculated using our non-local fracture model (Wang and Sharma 2017;Wang et al 2017), as shown in **Fig.22**. "Multilayer Case 1" is the most common scenario where height recession can happen during the fracture closure process where fracture closes first within formation barriers (e.g., Layer 1 and Layer 3) where stresses are higher. Compared to the single layer case in Fig.22, height recession (i.e., Multilayer Case 1) leads to an earlier and steeper increase in fracture stiffness. The fracture stiffness evolution of the single layer case and "Multilayer Case 2" are similar, because they have the same average stress.

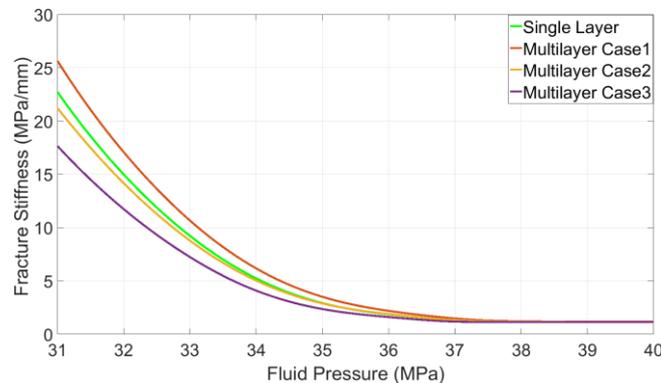

**Fig.22 Stiffness evolution for single and multilayer cases**

**Fig.23** shows the fracturing pressure and its derivatives for single and multilayer cases on G-function and square root of time plots. It can be observed that the trend of pressure derivatives generally reflects the evolution of fracture stiffness. The case of height recession (i.e., Multilayer Case 1) leads to earlier and steeper increase in pressure derivatives on both G-function and square root of time plots.

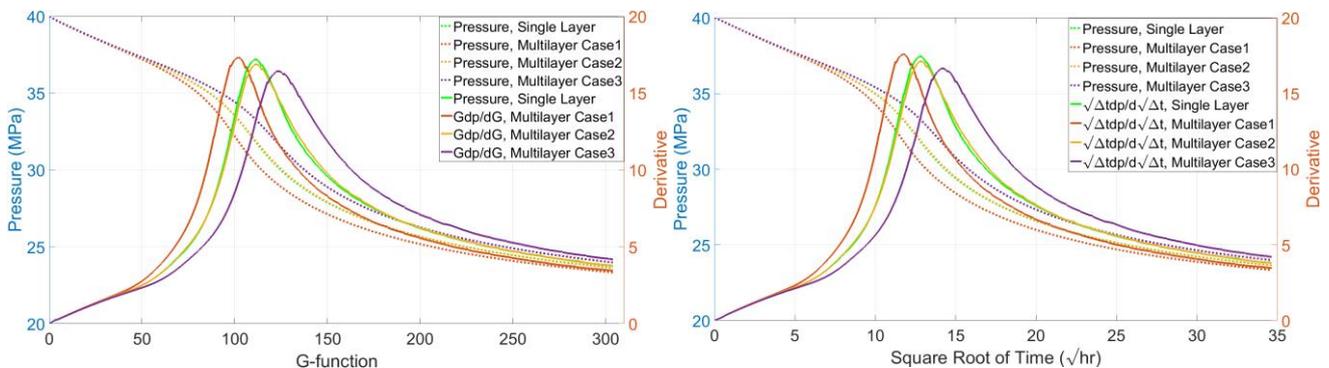

**Fig.23 Pressure decline response for single and multilayer cases**

A comparison of different methods of estimating minimum in-situ stress for the single and multilayer cases is presented in **Table 4**. Again, we can oberve that the tangent line method underestimates while the compliance method overestimates the minimum in-situ stress (i.e., 35 MPa in the target layer). Our "variable compliance method" gives the least errors for all cases.

| Scenarios | Tangent Line Method | | Compliance Method | | Variable Compliance Method | |
|---|---|---|---|---|---|---|
| | Estimated $\sigma_{hmin}$ (MPa) | Absolute Error (MPa) | Estimated $\sigma_{hmin}$ (MPa) | Absolute Error (MPa) | Estimated $\sigma_{hmin}$ (MPa) | Absolute Error (MPa) |
| Single Layer | 32.6 | -2.4 | 37.3 | 2.3 | 35.4 | 0.4 |
| Multilayer Case1 | 32.4 | -2.6 | 37.8 | 2.8 | 35.8 | 0.8 |
| Multilayer Case2 | 32.5 | -2.5 | 37.3 | 2.3 | 35.4 | 0.4 |
| Multilayer Case3 | 32.1 | -2.9 | 37.2 | 2.2 | 35.2 | 0.2 |

**Table 4-Comparison of different methods of estimating minimum in-situ stress for single and multilayer cases**

Based on the disscussion by McClure et al. (2016), the "compliance method" gives an estimate of the mecanical closure pressure, and this closure pressure can be 1-2 MPa above the actual minimum in-situ stress, depending on fracture geometry and contact width $w_0$. Our anlaysis in Table 3 shows that the difference between the infered closure pressure and input minimum in-situ stress that is constrained within 1-2 MPa is only limited to a few cases where fracture height is large or contact width $w_0$ is small. For a typical DFIT, the injecton volume is normally only a few cubic meters and the resulting fracture height or radius is often limited. In addition, depending on rock mineralogy and heterogeneity, the height of asperities can range from hundreds to thousands of micrometers (Bhide et al. 2014; Fredd et al. 2000; Sakaguchi et al. 2006; Wells and Davatzes 2015; Warpinski et al. 1993; Zou et al. 2015), and, as shown by the field tiltmeter survey, the residual fracture width can be as large as 20-30% of the maximum fracture width during pumping (Warpinski et al. 2002). Remember $w_0$ is determined by the tallest asperities, so it is challenging to calibrate the minimum in-situ stress using the "compliance method". In addtion, if the fracture has penetetrated into barrier layers with higher minimum pricinpal stress (e.g., Multilayer Case1 in Table 4), the fracture walls will come into contact at a higher pressure within the barrier layers and increase the overall fracture stiffness, so the "compliance method" can furthur overstimate the minimun in-situ stress. However, in all simulated cases the presented "variable compliance method" provides a much more consistent estimation of minimum in-situ stress. Most importantly, the results are insensitive to input parameters, which significantly reduces the uncertainties associated with the "compliance method".

### 3.2.3 Radial Fracture Geometry

Next, we investigate the pressure decline response for radial fracture geometry, using parameters from Table 2 and assuming $w_0$ is 2 mm, $\sigma_{ref}$ is 5 MPa, fracture radius is 10 m and reservoir permeability is 0.0005 md as the Base Case. Fracturing pressure decline response for different $w_0$ and are shown in **Fig.24**. The results and trends are similar to those obtained for the PKN geometry, a small contact width leads to steeper changes in pressure decline rate and the slope of pressure derivative, and increases the peak value of pressure derivative.

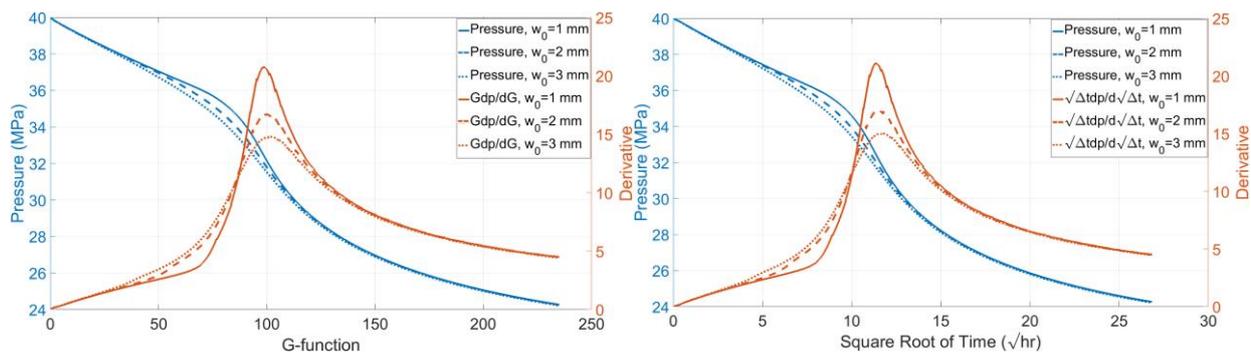

**Fig.24 Pressure decline response for different $w_0$ with a radial geometry**

**Fig.25** shows the pressure decline response on G-function and the square root of time plots for different contact reference stress. As in the PKN cases, a larger value of contact reference stress increases the pressure decline rate after the fracturing pressure drops to a certain level and reduces the peak value of pressure derivatives.

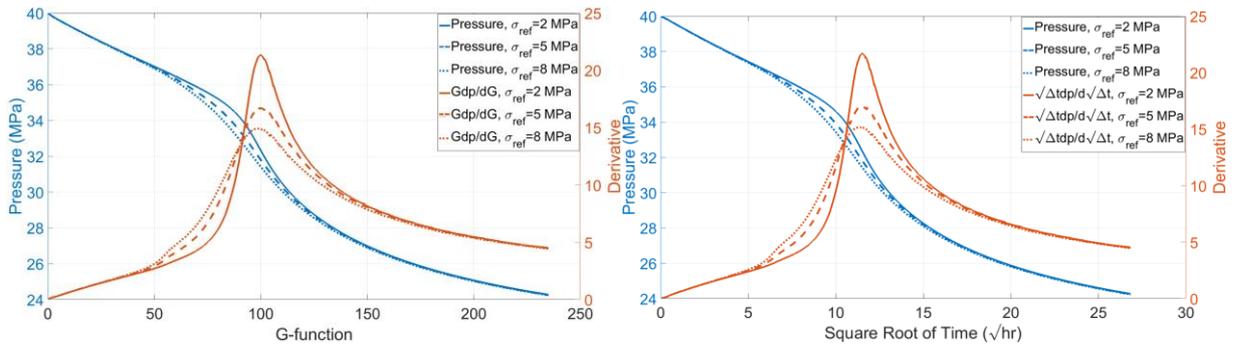

**Fig.25 Pressure decline response for different $\sigma_{\text{ref}}$ with a radial geometry**

**Fig.26** shows the corresponding pressure decline response on G-function and the square root of time plots. A larger fracture radius leads to lower fracture stiffness and this, in turn, results in a lower pressure decline rate.

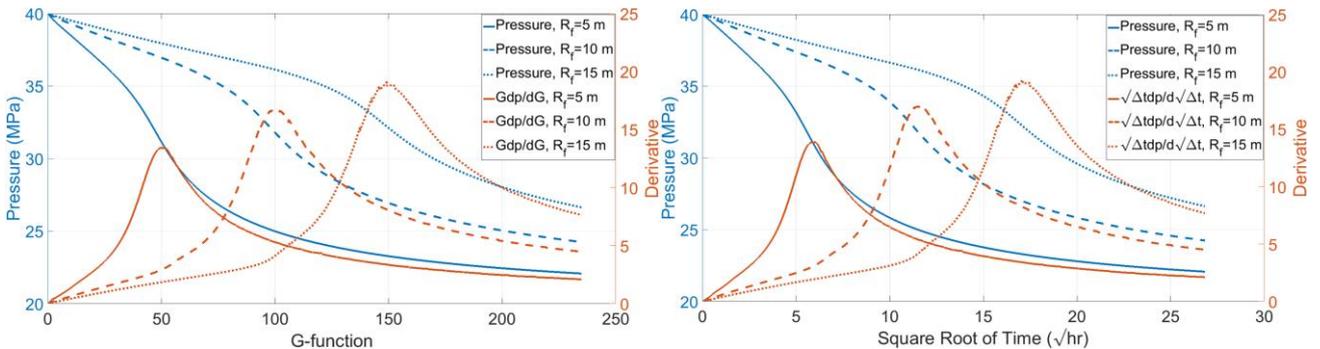

**Fig.26 Pressure decline response for different fracture radii with a radial geometry.**

**Fig.27** shows the fracturing pressure and its derivatives for different reservoir permeability on G-function and the square root of time plots. As expected, pressure declines more rapidly when the reservoir permeability is large. At early time, the pressure decline rate increases because an increase in the fracture stiffness dominates changes in leak-off rate. At late time when the fracture becomes less compliant, the pressure decline rate decreases due to smaller and smaller differences between the fracture pressure and the initial reservoir pressure. Eventually, the leak-off rate will drop to zero as the fracture pressure becomes equal to the reservoir pressure.

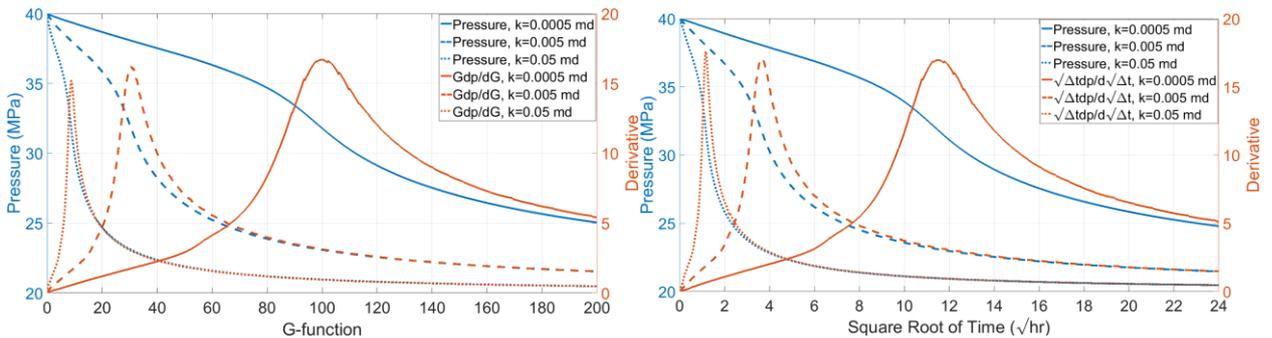

**Fig.27 Pressure decline response for different reservoir permeability with a radial geometry**

**Fig.28** shows the pressure decline response for different wellbore volumes on G-function and the square root of time plots. Compared to Fig. 19, it can be observed that the WBS effect has stronger influence on pressure decline trend in our radial fracture geometry for the same wellbore volume. This is because the fracture volume and fracture surface are smaller than that in previous case of PKN fracture geometry, which intensifies the impact of WBS effect.

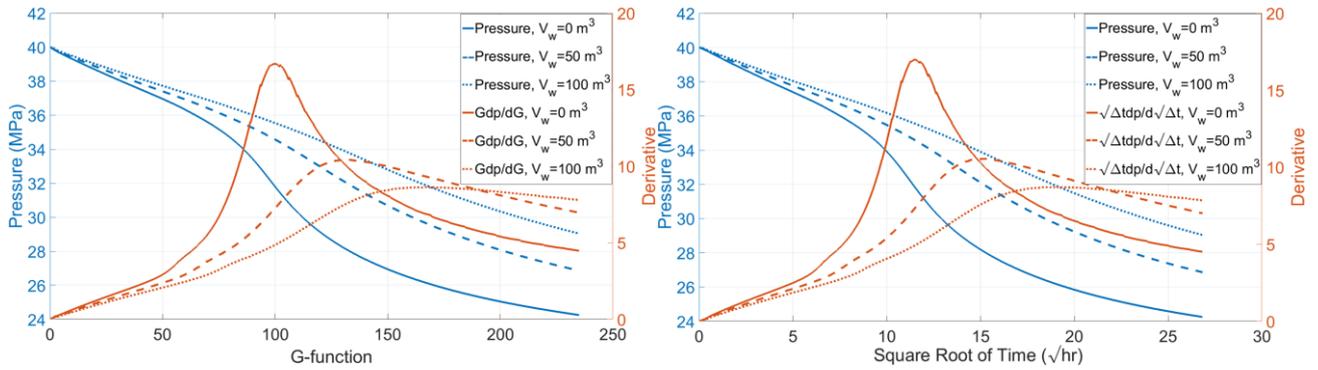

**Fig.28 Pressure decline response for different wellbore volume with a radial geometry**

**Fig.29** shows the fracturing pressure and its derivatives for different initial reservoir pressure on G-function and square root of time plots. Similar to Fig.20, pressure declines rapidly in depleted reservoirs and declines gradually with much lower peak values of pressure derivatives in over-pressurized reservoirs. Also the pressure decline trend resembles a "normal-leak-off behavior" when pore pressure is abnormally high.

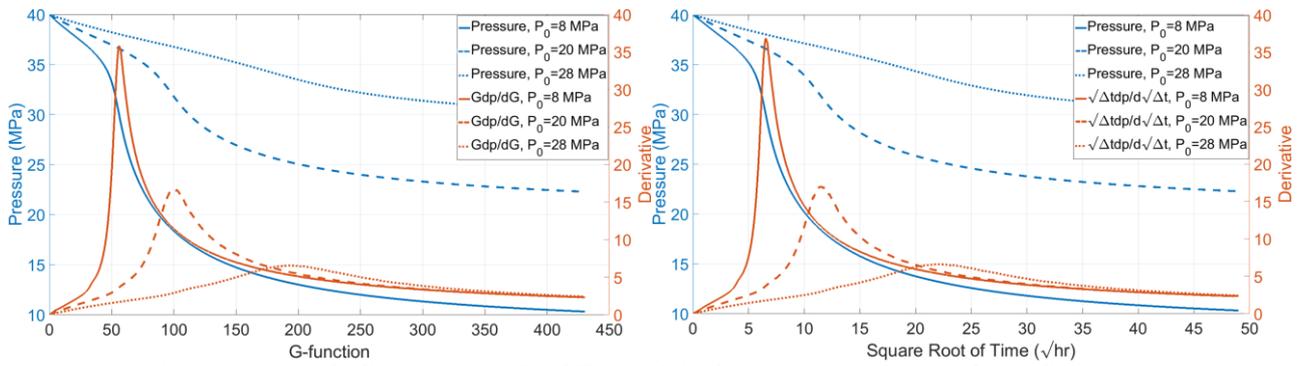

**Fig.29 Pressure decline response for different initial reservoir pressure with a radial geometry**

**Table 5** shows a comparison of the different methods of estimating minimum in-situ stress (closure pressure is interpreted as the minimum in-situ stress) for the above synthetic cases for a radial fracture geometry. Again, the results indicate that the "tangent line method" always underestimates minimum in-situ stress while the "compliance method" overestimates the minimum in-situ stress (if the closure stress is assumed to be equal to the minimum in-situ stress) if no correction is applied (except for the so-called "normal leak-off behavior" where the changes in fracture stiffness/compliance are undetectable). The estimated minimum in-situ stress using our proposed "variable compliance method" give the smallest errors in all simulated cases.

| | Scenarios | Tangent Line Method | | Compliance Method | | Variable Compliance Method | |
|---|---|---|---|---|---|---|---|
| | | Estimated $\sigma_{hmin}$ (MPa) | Absolute Error (MPa) | Estimated $\sigma_{hmin}$ (MPa) | Absolute Error (MPa) | Estimated $\sigma_{hmin}$ (MPa) | Absolute Error (MPa) |
| | Base Case | 32.3 | -2.7 | 37.3 | 2.3 | 35.3 | 0.3 |
| | $w_0$=1 mm | 32.7 | -2.3 | 36.7 | 1.7 | 35.1 | 0.1 |
| | $w_0$=3 mm | 31.7 | -3.3 | 38.2 | 3.2 | 35.1 | 0.1 |
| | $\sigma_{ref}$=2 MPa | 32.4 | -2.6 | 36.9 | 1.9 | 35.3 | 0.3 |
| | $\sigma_{ref}$=8 MPa | 32.2 | -2.8 | 37.3 | 2.3 | 35.2 | 0.2 |
| | $R_f$=5 m | 31.3 | -3.7 | 39.3 | 4.3 | 35.3 | 0.3 |
| Modified Base Case | $R_f$=15 m | 32.5 | -2.5 | 36.8 | 1.8 | 35.2 | 0.2 |
| | k=0.005 md | 32 | -3 | 37.4 | 2.4 | 35.1 | 0.1 |
| | k=0.05 md | 31.8 | -3.2 | 37.5 | 2.5 | 35.3 | 0.3 |
| | $V_w$=50 $m^3$ | 32.7 | -2.3 | 37.1 | 2.1 | 35.2 | 0.2 |
| | $V_w$=100 $m^3$ | 32.8 | -2.2 | 37.1 | 2.1 | 35.3 | 0.3 |
| | $P_0$=8 MPa | 30.5 | -4.5 | 37.0 | 2 | 35.2 | 0.2 |
| | $P_0$=28 MPa | 34.5 | -0.5 | NA | NA | NA | NA |

**Table 5-Comparison of different methods of estimating minimum in-situ stress for radial fracture geometry.**

From the analysis presented in this section, we can conclude that the pressure decline response during DFIT is affected by many parameters, such as fracture geometry, reservoir properties, fracture surface roughness, etc. As a result any history match DFIT data can be non-unique if we don't have good knowledge (from other independent analysis and measurements) of the constraints on these parameters. Nevertheless, our proposed "variable compliance method" gives the most reliable estimation of minimum in-situ stress regardless of these parameters, so the uncertainties associated with fracture geometry, surface roughness and reservoir properties do not impact our estimation of minimum in-situ stress.

## 4. Field Data Interpretation

### 4.1 Field Case 1

The first field case to be analyzed comes from a horizontal well-A drilled through a shale formation. The measured depth is around 5500 m and a diagnostic fracture injection test is conducted at the toe of the horizontal wellbore, with 2.35 m³ of water injection in 3 minutes, then the well was shut-in for 27 days. **Fig.30** shows the pressure decline response on G-function and the square root of time plots. As in the previous synthetic cases, even though G-function and the square root of time plots give exactly the same quantitative information, it is more intuitive to infer the shut-in time on the square root of time plot, while the shut-in time is implicit in the value of G-function, as reflected by Eq.(3)~Eq.(5). This field case also exhibits linear pressure derivatives and low peak value of pressure derivative with a long extended tail after it reaches the peak, we can infer that the impact of wellbore storage should be substantial and this reservoir is potentially over-pressurized.

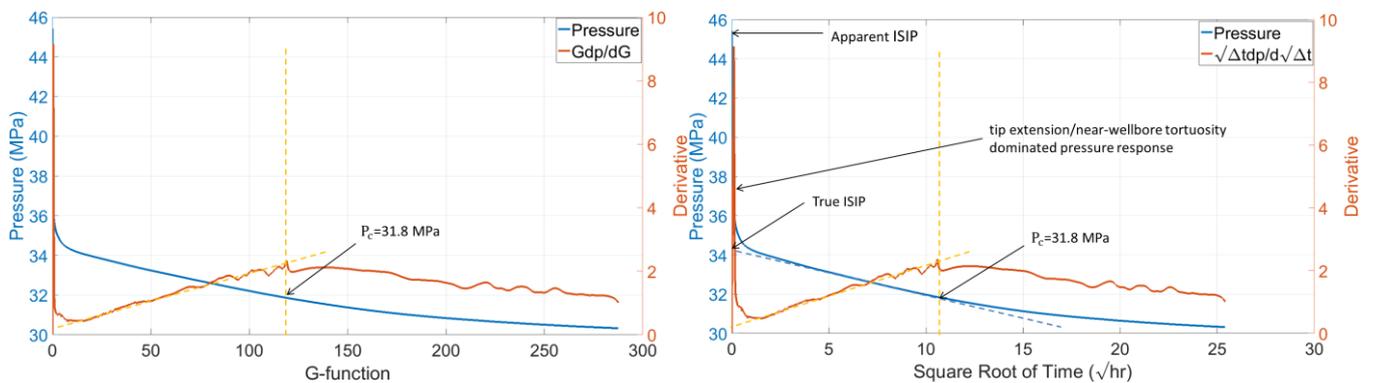

**Fig.30 Pressure decline response from Well-A on G-function and square root of time plots**

From the signature of pressure derivatives, the changes in fracture stiffness/compliance are undetectable, so the minimum in-situ stress is picked at 31.8 MPa as the "tangent line method" suggests. A closer look at the square root of time plot, shows that the pressure drops significantly during the first one hour of shut-in. By examining the pressure derivative curve, whose extrapolated value is not zero when $\Delta t = 0$, we can identify that the excessive pressure drop at the beginning is caused by tip extension or near-wellbore tortuosity. The apparent ISIP is 45.5 MPa while the true ISIP is 34.5 MPa. Remember net pressure is the pressure difference between the fluid pressure inside the fracture and closure pressure. In reality, the pressure measured at the wellhead can be much higher than expected during fracture propagation during the early time of shut-in period, because significant pressure drop can happen in the near-wellbore region and along the horizontal wellbore, due to friction and tortuous/complex fracture path that initiated from perforation clusters. So pressure data from this period cannot be used to infer fracturing parameters and reservoir properties. If a rate step down test can be performed before complete shut-in, the relationship between early-time excessive pressure drop and flow rate can be quantified.

Log-log plots of pressure derivatives are often used to determine flow regimes for after closure analysis and assist the interpretation of closure stress (Barree et al. 2019; Hawkes et al. 2012; Soliman et al. 2015). Even though Well-A was shut-in for nearly four weeks, pseudo-radial flow is still absent, as shown in **Fig. 31**. This is a clear evidence of extremely low formation permeability. Without pseudo-radial flow, formation pore pressure and permeability cannot be determined independently with enough confidence using just after closure analysis alone. One has to analyze the whole spectrum of DFIT data to reach a consistent interpretation. Using late time data from the linear flow regime, and plotting $P_f$ vs $1/\sqrt{\Delta t}$ on Cartesian coordinates, the intercept on the y-axis when $\Delta t \to \infty$ is the estimated initial pore pressure, which is 29.1 MPa. This indicates it is indeed an over-pressurized reservoir. From geological and petrophysical studies, it is known that the thickness of the target formation is 24.4 m with an average porosity of 0.07 and in-situ fluid viscosity of 0.257 cp, The Young's modulus is 38.9 GPa, the Poisson's ratio is 0.2 and formation total compressibility is 3e-3 MPa⁻¹. Hydraulic fracture modeling indicates that the fracture is well contained within the target formation with penny-shaped fracture geometry, and the fracture radius is roughly 12m. Based on this information, the pressure decline response can be matched globally using our fully coupled DFIT model (Eq.(6)-Eq.(16)) with pressure dependent fracture stiffness obtained from our non-local fracture closure model (Wang and Sharma 2017; Wang et al. 2017), from the end of tip extension/near-wellbore tortuosity dominated period to the end of the test.

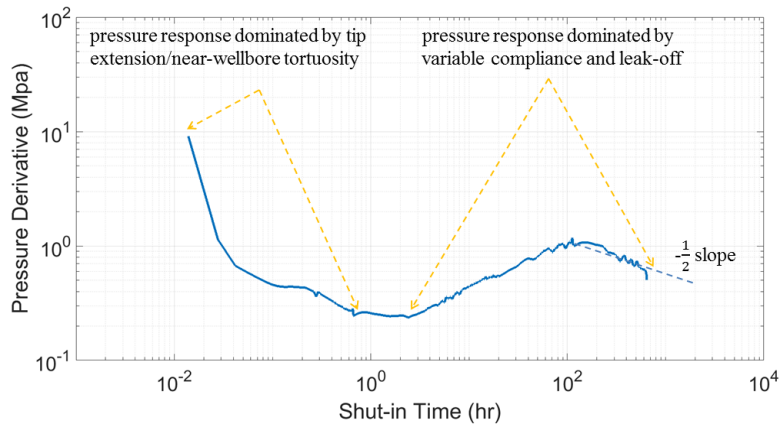

**Fig.31 Log-log pressure derivative plot of Well-A**

**Fig.32** shows the pressure decline response predicted by our fully coupled DFIT model and field data on G-function and the square root of time plots. The results indicate that our simulated pressure matches extremely well with the field data for the entire duration of the test, excluding the first hour of shut-in. Our matched reservoir permeability is 220 nd, which is within the range of other petrophysical measurements. The matched contact width and contact reference stress are 0.7 mm and 3 MPa, respectively, and **Fig.33** shows the corresponding fracture and fracture-wellbore system stiffness, based on the matched fracture geometry, wellbore volume and contact parameters. Because the DFIT was conducted at the toe of a horizontal well, it is no surprise that the wellbore storage effect is significant, considering such a large contrast between the wellbore and fracture volume. We can also observe that the fracture stiffness starts to increase noticeably when pressure drops to 34 MPa, over 2 MPa above the minimum in-situ stress. It is apparent that the so-called "normal leak-off behavior" is just a coincidence when the effect of pressure dependent leak-off at the fracture surface and variable fracture compliance counterbalance each other during the fracture closure process.

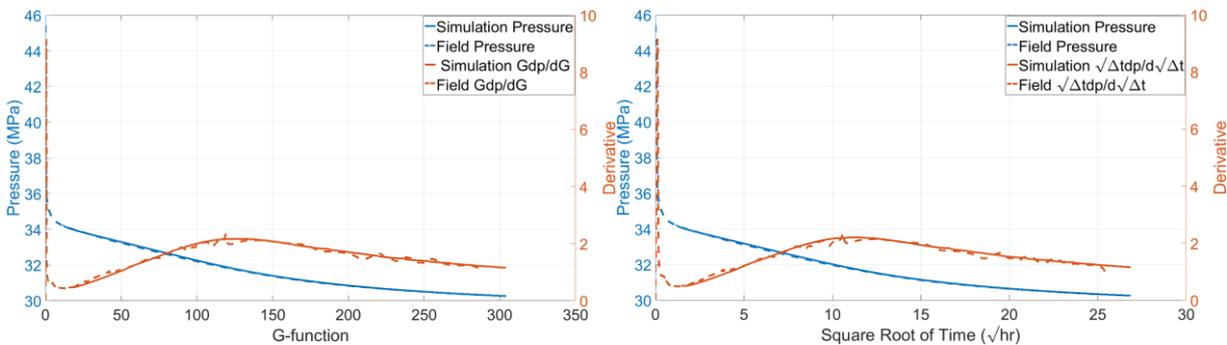

**Fig.32 Matched pressure decline response for Well-A on G-function and square root of time plots**

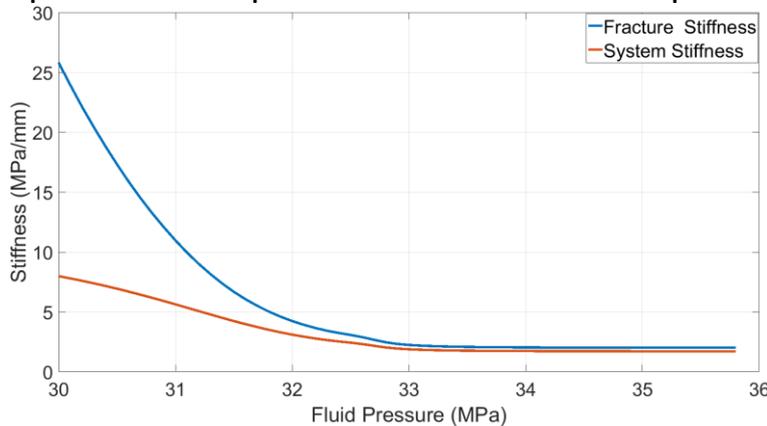

**Fig.33 Matched fracture and fracture-wellbore system stiffness for Well-A**

### 4.2 Field Case 2

The second field case to be analyzed comes from a vertical well-B drilled through a shale formation. The total wellbore length is around 2000 m and a diagnostic fracture injection test is conducted with 4.7 m$^3$ of water injection in 6 minutes, then the well was shut-in for 11 days. **Fig.34** shows the pressure decline response on G-function and the square root of time plots. If we pick the fracture closure pressure using the "tangent line method", both G-function and the square root of time plots give the same minimum in-situ stress of 50.4 MPa. One can also notice that there is a slight "bump" in the pressure derivatives soon after shut-in. For shale formations, this early-time accelerated pressure drop is not likely to be caused by PDL, but attributed to tip extension and near wellbore tortuosity (continued after-flow of fluid from wellbore to fracture after shut-in). It should be observed that in the early-time pressure transient response, the pressure derivative starts to deviate upwards at around 55 MPa. Using "variable compliance method" through averaging the G-function or the square root of time that corresponding to a pressure of 50.4 MPa and 55 MPa, respectively, we can obtain the estimated minimum in-situ stress of 52.8 MPa by picking the pressure that corresponds to the averaged G-function or square root of time.

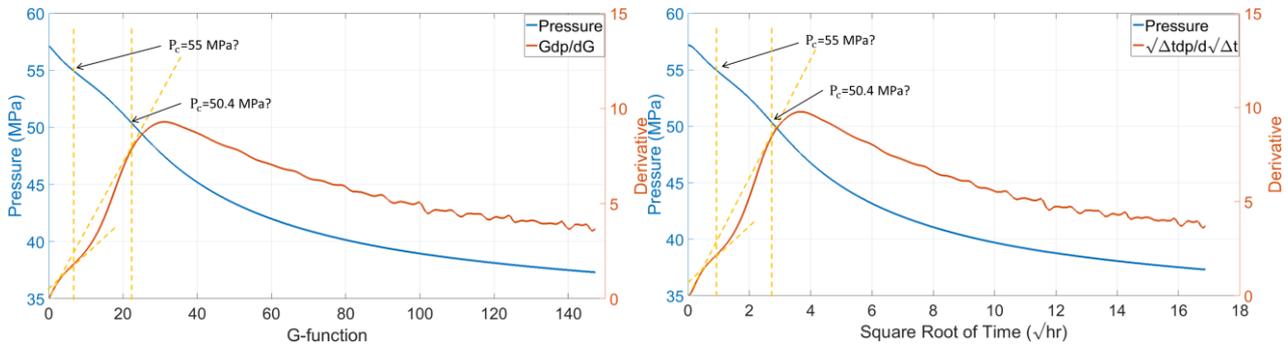

**Fig.34 Pressure decline response from Well-B on G-function and square root of time plots**

From **Fig. 35**, we can observe that only linear flow emerged after closure. Using late time data from the linear flow regime, the estimated initial pore pressure is 33.7MPa. Geological and petrophysical studies indicate that the thickness of the target formation is 5 m with an average porosity of 0.03 and in-situ fluid viscosity of 0.28 cp. The Young's modulus is 39.5 GPa, the Poisson's ratio is 0.25 and formation total compressibility is $1.9e-3$ MPa$^{-1}$. Hydraulic fracture modeling shows that the fracture is well contained within the target formation with roughly 200 m fracture half length. Based on this information and assuming PKN geometry, the pressure decline response can be matched globally using our global DFIT model by adjusting reservoir permeability and contact parameters.

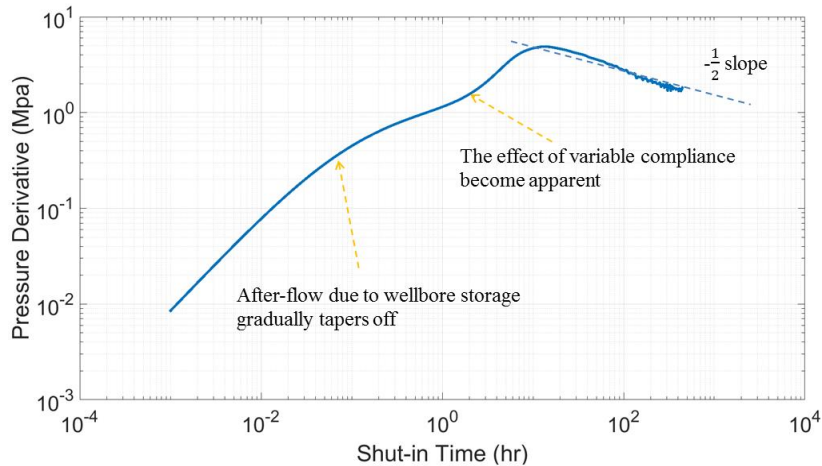

**Fig.35 Log-log pressure derivative plot of Well-B**

**Fig.36** shows the predicted pressure decline response and field data on G-function and the square root of time plots. The results indicate that our simulated pressure agrees well with field data and it even captures the downward curve of pressure derivatives before closure. Traditionally, this phenomenon is interpreted as "height recession" or "transverse storage", where the top and bottom barrier formations close first and squeeze fluid into the target formation during the closure process. This can be regarded as a special case of "variable compliance", as demonstrated in Fig.23.. Because fractures always close from their edges to the center on rough walls, this constantly alters fracture compliance/stiffness. This variable fracture compliance coupled with fracture pressure dependent leak-off dictates our pressure decline trend for a given formation. In this field case, our matched reservoir permeability is 210 nd, and the matched contact width and contact reference stress are 1.2 mm and 1.1 MPa, respectively.

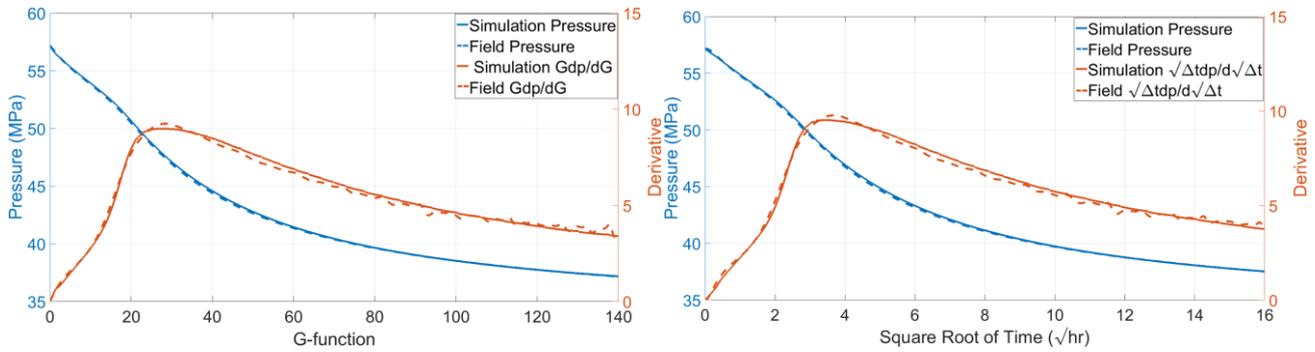

**Fig.36 Matched pressure decline response for Well-B on G-function and square root of time plots**

**Fig.37** shows the corresponding fracture and fracture-wellbore system stiffness as a function of fluid pressure inside fracture. It can be observed that both fracture and system stiffness increase gradually as pressure declines. Even though this test was conducted in a vertical well with relative moderate wellbore volume, the wellbore storage effect on the system stiffness is still significant. Re-examining Eq.(16), we can see that the relative influence of wellbore storage not only depends upon the ratio of wellbore to fracture volume, but it also depends on fracture stiffness itself. For the same wellbore volume, the higher the fracture stiffness, the less compressible the fracture is, so the compressed fluid in the wellbore plays a bigger role to compensate leak-off and pressure drop. In this field case, with fracture height contained in 5 m, the fracture stiffness should be at least 5.36 MPa/mm (From Table 1) even without considering the contact of fracture surfaces. If the fracture geometry and reservoir properties can be constrained with enough confidence, the estimated contact parameters and fracture stiffness evolution can be used to reconstruct the dynamic fracture closure process and the evolution of fracture width distribution, thus, the un-propped fracture conductivity and its pressure/stress dependence can be inferred.

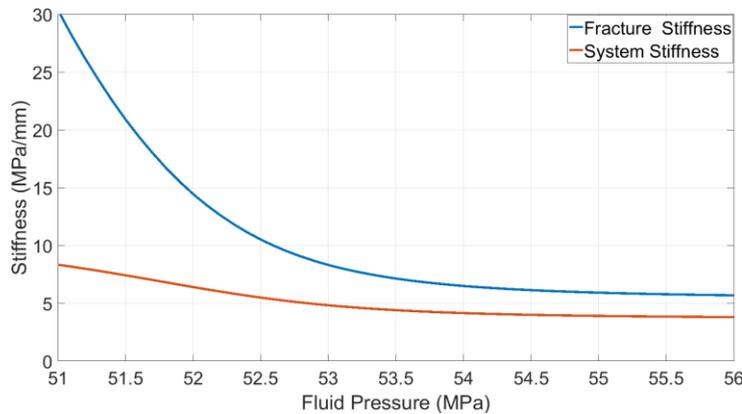

**Fig.37 Matched fracture and fracture-wellbore system stiffness for Well-B**

## 5. Conclusions and Discussion

Since the introduction of Nolte's pioneering work, diagnostic fracture injection test (DFIT) has evolved into a commonly used technique and are generally accepted as a reliable way to obtain fracturing parameters (e.g., minimum in-situ stress, leak-off behavior) and reservoir properties (e.g., initial pore pressure and representative formation permeability). These key parameters are needed to run fracture models, post-fracture production prediction and economical evaluations. In this article, we first reviewed Nolte's G-function model and the associated assumptions (adopted by many subsequent works), and derived the equations that underpin the interpretation of the square root of time plot. We demonstrated that the square root of time plot gives the same quantitative information obtained from the G-function plot.

A new global DFIT model is proposed, which for the first time, has the capability to predict pressure decline response across the entire duration of a DFIT test using a single, coherent mathematic framework. We show how each factor impacts the pressure decline response and the effects of previous overlooked coupled mechanisms are examined and discussed, along with synthetic and field case studies. The content of this article promotes our conceptual understanding of pressure transient and coupled behaviors during DIFT and enables us to revisit DFIT analysis from a new perspective. Conclusions reached from the analysis presented in this paper include the following.

1. Carter's leak-off, (leak-off rate is proportional to $1/\sqrt{\Delta t}$ ), is applicable during stable fracture propagation, when net pressure is roughly constant and injection time is limited. However, during the fracture closure peroid, the fluid pressure in the fracture declines significantly, and the leak-off rate becomes fracture pressure dependent and ceases to be proportional to $1/\sqrt{\Delta t}$ .

2. The so-called "normal leak-off", with a straight line of pressure and its derivatives on G-function and the square root of time plots, is not "normal" at all. This is sometimes observed in data sets when mechanisms that accelerate pressure decline (e.g., increase in fracture stiffness during closure) offset those that decelerate pressure decline (e.g., pressure dependent leak-off at the fracture surface and wellbore storage effects). So any "before closure" analysis based on this conventional "normal leak-off" assumption to estimate fracture geometry and formation permeability should be re-examined.

3. Hydraulic fractures do not close as two smooth parallel plates, but rather close progressively from edges to the center on asperities and roughness surfaces, as confirmed by our modeling results and field tiltmeter measurements. The fracture stiffness increases (compliance decreases) gradually as more and more fracture surfaces come into contact, even before the fracture pressure drops to the minimum in-situ stress.

4. It is shown that the pressure decline response during fracture closure is governed by a PDE-ODE system, where the pressure at the fracture surface should be solved simultaneously with the pressure inside the formation. Previous DFIT models assumed a constant fracture compliance/stiffness and Carter's leak-off oversimplified and decoupled this system, which prevents them from predicting pressure decline across the entire duration of the DFIT test.

5. The conventional "tangent line method" underestimates the minimum in-situ stress, especially in depleted reservoirs. The "compliance method" picks the mechanical closure pressure but overestimates the minimum in-situ stress (if the two are equated), especially when height recession occurs. Height recession (or transverse storage) is a just special type of variable compliance phenomenon.

6. The "variable compliance method" proposed in this study suggests that picking an average of the dimensionless G-time (or the square root of time) of the two methods provides a much more robust and reliable way of estimating the minimum in-situ stress when concave pressure derivative exist. The uncertainties associated with fracture geometry, surface roughness and reservoir properties do not impact our estimation of minimum in-situ stress using "variable compliance method". Results using this method match well with field data.

7. Fracture closure on asperities and rough fracture surfaces is a gradual process starting at the fracture tip. The model presented here can be used to define the relationship between fracture width and contact stress using DFIT pressure data. If other matching parameters can be constrained with enough certainty, this information can be used to infer unpropped fracture conductivity as a function of effective stress. This adds tremendous value to DFIT analysis for completion design and estimates of well productivity and EUR.

## Nomenclature

| | |
|---|---|
| $a$ | = Half fracture height (PKN model) or half fracture length (KGD model), L, $m$ |
| $A_f$ | = Fracture surface area (one face of one wing), $L^2$, $m^2$ |
| $c_t$ | = Formation total compressibility, $Lt^2/m$, 1/Pa |
| $c_w$ | = Water compressibility, $Lt^2/m$, 1/Pa |
| $C_L$ | = Carter's leak-off coefficient, $L/\sqrt{t}$, $m/\sqrt{s}$ |
| $E$ | = Young's modulus, $m/Lt^2$, $Pa$ |
| $E'$ | = Plane strain Young's modulus, $m/Lt^2$, $Pa$ |
| $g(\Delta t_D)$ | = Dimensionless g-function of time |
| $G(\Delta t_D)$ | = Dimensionless G-function of time |
| $h_f$ | = Fracture height, L, $m$ |
| $ISIP$ | = Instant shut-in pressure, $m/Lt^2$, $Pa$ |
| $k$ | = Formation permeability, $L^2$, $m^2$ |
| $P$ | = Pressure, $m/Lt^2$, $Pa$ |
| $P_f$ | = Fracturing pressure, $m/Lt^2$, $Pa$ |
| $P_{net}$ | = Fracturing net pressure, $m/Lt^2$, $Pa$ |
| $P_0$ | = Initial reservoir pressure, $m/Lt^2$, $Pa$ |
| $q_f$ | = Leak-off rate (one wing), $L^3/t$, $m^3/s$ |
| $r_D$ | = Normalized radius |
| $r_p$ | = Productive surface ratio, which is the ratio of fracture surface area that is subject to leak-off to the total fracture surface area. |
| $r_f$ | = Local fracture radius, $L, m$ |
| $R_f$ | = Fracture radius, $L, m$ |
| $S_f$ | = Fracture stiffness, which is the reciprocal of fracture compliance, $m/L^2t^2$, $Pa/m$ |
| $S_s$ | = Fracture-wellbore system stiffness, $m/L^2t^2$, $Pa/m$ |
| $t$ | = Generic time, $t$, $s$ |
| $t_D$ | = Dimensionless time |
| $t_e$ | = Generic time at the end of pumping, $t$, $s$ |
| $t_p$ | = Pumping time, $t$, $s$ |
| $\Delta t$ | = Total shut-in time, $t$, $s$ |
| $\Delta t_D$ | = Dimensionless shut-in time, $t$, $s$ |
| $x_f$ | = Fracture half-length, $L$, $m$ |
| $U_f$ | = Flux velocity through fracture surface with constant fracturing pressure, $L/t$, m/s |
| $V_f$ | = Fracture volume (one wing), $L^3$, $m^3$ |
| $V_w$ | = Half of wellbore volume, $L^3$, $m^3$ |
| $w_0$ | =Contact width, L, $m$ |
| $w_f$ | =Local fracture width, L, $m$ |
| $\bar{w}_f$ | =Average fracture width, L, $m$ |
| $\mu_f$ | = Fluid viscosity, m/Lt, Pa s |
| $\nu$ | = Poisson's ratio |
| $\sigma_c$ | = Contact stress, m/L$t^2$, $Pa$ |
| $\sigma_{hmin}$ | = Minimum in-situ stress, m/L$t^2$, $Pa$ |
| $\sigma_{ref}$ | = Contact reference stress, m/L$t^2$, $Pa$ |
| $\phi$ | = Formation porosity |
| Subscript | |
| n | = Variables at n[th] time interval |
| j | = Variables at j[th] time interval |
| i | = Variables at j[th] time interval |

## Appendix : Solutions of Pressure Decline under Fracture Pressure Dependent Leak-off

In order to overcome the limitations of Nolte's G-function model and circumvent the use of a constant leak-off coefficient, Mayerhofer et al. (1995) presented a model that describes unsteady-state fluid flow from fractures of varying area into the formation, with the filter cake considered as a time and rate dependent skin effect, allowing for superposition of the injection history, filter cake deposition and associated rate convolution. Valko ´ and Economides (1999) simplified Mayerhofer et al. (1995) model by assuming constant leak-off rate during fracture propagation and constant fracture area during closure. A special before closure analysis (BCA) plot with a straight line can be constructed to determine reservoir permeability and fracture face resistance, without acquiring the initial guess of permeability and fracture area. However, the process to generate the special BCA plot and predict fall-off pressure is complicated and cumbersome, prone to mistakes that can lead to dimension inconsistency (Soliman et al. 2013). In addition, their approach used a constant-rate solution of an infinite-conductivity well introduced by Gringarten et al. (1974). The results obtained by use of the constant-rate solution are questionable to account for the fracture pressure dependent leak-off behavior. In this appendix, we derive the solutions of pressure decline under fracture pressure dependent leak-off using superposition based on constant-pressure solution.

The basic principle of superposition is that the response of a system to a number of perturbations is equal to the sum of the responses to each of the perturbation as if they were present by themselves if the responses are linearly related to the perturbations. From Eq.(18), we can calculate the leak-off rate under a constant fracture pressure $P_f$:

$$q_f = 2A_f(P_f - P_0)\sqrt{\frac{k\phi c_t}{\pi \mu_f \Delta t}} \qquad (A.1)$$

Divide the shut-in time $\Delta t$ into n time steps, and the leak-off rate at the $n^{th}$ time step can be determined based on superposition:

$$q_{f,n} = 2A_f\sqrt{\frac{k\phi c_t}{\pi \mu_f}} \sum_{j=1}^{n} \frac{P_{f,j} - P_{f,j-1}}{\sqrt{\Delta t_n - \Delta t_{j-1}}} \qquad (A.2)$$

The pressure difference $P_{f,j} - P_{f,j-1}$ is the pressure in the fracture at time step j minus the pressure in the formation at the fracture-formation interface, which equals to the pressure in the fracture at the previous time step. Combine Eq.(A.2) with Eq.(10) via material balance, we can get:

$$-\frac{1}{S_f}\frac{dP_{f,n}}{d\Delta t_n} = 2\sqrt{\frac{k\phi c_t}{\pi\mu_f}}\sum_{j=1}^{n}\frac{P_{f,j}-P_{f,j-1}}{\sqrt{\Delta t_n - \Delta t_{j-1}}} \tag{A.3}$$

let $P_{f,0} = P_0$, $P_{f,1} = \text{ISIP}$, $\Delta t_1 = 0$, then for $n \geq 2$, integrate Eq.(A.3) we can obtain :

$$P_{f,n} = \text{ISIP} - 4S_f\sqrt{\frac{k\phi c_t}{\pi\mu_f}}\sum_{n=1}^{n-1}\sum_{j=1}^{n-1}(P_{f,j}-P_{f,j-1})(\sqrt{\Delta t_n - \Delta t_{j-1}} - \sqrt{\Delta t_{n-1} - \Delta t_{j-1}}) \tag{A.4}$$

Eq.(A.4) can be solved explicitly with marching time interval, for example, when $n = 2$

$$P_f(\Delta t_2) = \text{ISIP} - 4S_f(\text{ISIP} - P_0)\sqrt{\frac{k\phi c_t \Delta t_2}{\pi\mu_f}} \tag{A.5}$$

On closer observation of Eq.(A.5), we can realize that it is exactly Carter's leak-off solution to Eq.(19), because during the first time step, no superposition is needed. The fracture pressure dependent leak-off coefficient can be recovered as:

$$C_L = (P_f - P_0)\sqrt{\frac{k\phi c_t}{\pi\mu_f}} \tag{A.6}$$

## A Discussion of the Theoretical Basis of Closure Stress Estimation Using Variable Compliance Method

We know that the pressure decline response during fracture closure is governed by a PDE-ODE system, where the pressure at the fracture surface should be solved simultaneously with the pressure inside the formation. Previous DFIT models assumed a constant fracture compliance/stiffness and Carter's leak-off oversimplified and decoupled this system, which prevents them from predicting pressure decline across the entire duration of the DFIT test and leads to erroneous estimation of closure stress. A recent study by van den Hoek (SPE181593) also reveals that the peak of pressure derivatives on G-function plot and the end of 3/2 slope on log-log only mark the end of fracture storage dominated flow, and has nothing to do with fracture closure stress. This study agrees well with Wang and Sharma (SPE187348) that there is not a district feature on pressure transient response that we can pick closure stress directly from pressure derivative curves, nevertheless, the "variable compliance method" that proposed by Wang and Sharma (SPE187348) that gives the most reliable estimation of closure stress, is not merely a coincidence.

Craig et al. (SPE187038) compared the closure stress estimation from "tangent line method (Barree et al. 2009)" and "compliance method (McClure et al. 2016)" against tiltmeter measurement (measures rock deformation with downhole). Their study shows that "tangent line method" picks closure stress at the point where tiltmeter measurement starts to deviates from a horizontal line (i.e. fracture stiffness is infinite or fracture compliance is zero), as shown in **Fig. B1**. The closure stress estimated by these two methods has 6 MPa difference. The reason for the underestimation of closure stress by "tangent line method" is that when fracturing pressure drops to closure stress, even though all fracture surfaces have already closed on asperities, the fracture compliance is not zero. Depending on surface roughness (i.e. contact parameters), the fracture still has some compliance (also reflected in Fig.1), which is not small enough to be considered as negligible. The reason for the overestimation of closure stress by "compliance method" is that fracture compliance starts to change when fracture begins closure on tip regions, while fracturing pressure still higher than the closure stress.

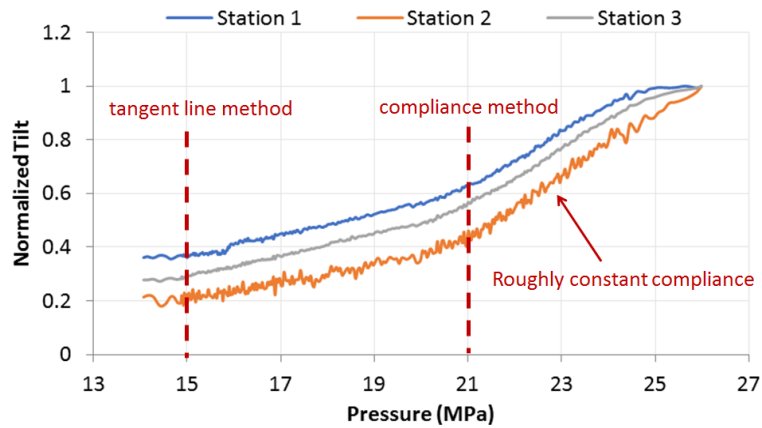

**Fig.B1 Normalized tiltmeter data from shut-in of 2B from the GRI/DOE M-site.**

Sensitivity analysis (SPE187348) shows that the "variable compliance method" using the G-function or square root of time based averaging approach to estimate closure stress is insensitive to fracture geometry, surface roughness and reservoir properties. However, they have not discussed the fundamental basis for the reliability of "variable compliance method" in their synthetic cases. Here we present an example to demonstrate the reason. **Fig.B2** shows the average fracture width (should be proportional to tiltmeter measurement) evolution for a PKN fracture geometry for different contact width $w_0$, assuming a Young's modulus of 20 GPa, Poisson's ratio of 0.25, $\sigma_{ref}$ of 5 MPa with 10 m fracture height, 18 MPa minimum in-situ stress. It can be noticed that as $w_0$ increases, the difference of estimated closure stress between "tangent line method" and "compliance method" also increases. For example, the difference is around 4.5 MPa when $w_0$=1mm while increases to 6 MPa when $w_0$=2 mm. This because when $w_0$ is larger, fracture is closes on tip regions at a higher pressure, which in turn, increase fracture stiffness (or decrease compliance). On the other hand, large $w_0$ makes the residual fracture more compressible, a lower fracturing pressure is needed to make the residual fracture compressibility negligible. However, to obtain reliable closure stress estimation, simply average the closure stress estimated by "tangent line method" and "compliance method" is not an appropriate solution, as reflected in Fig. B2. In addition, the leak-off also plays an important role in pressure decline response. The "variable compliance method" provides an alternative way to estimate closure stress that not only accounts for the difference between two end points (fracture compliance start to change and fracture compliance become small enough to be considered as negligible), but also compensates for the impact of leak-off, which reflected on G-function time or the square root of time. That's why the closure stress estimated by "variable compliance method" is reliable and can be used as an anchor point for further DFIT analysis.

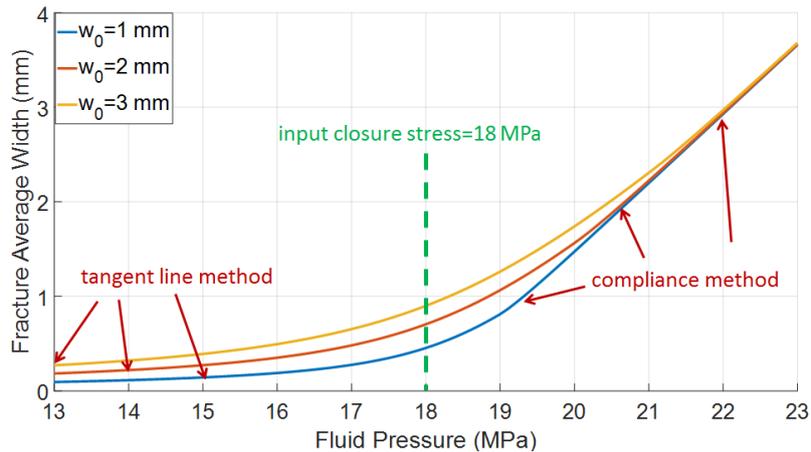

**Fig.B2 Fracture average width evolution for a PKN fracture geometry**